\documentclass{aa}

\usepackage{gensymb}

\usepackage{changepage}
\usepackage{amssymb}
\usepackage{graphicx}
\usepackage{natbib}
\bibpunct{(}{)}{;}{a}{}{,} 
\usepackage{bm}
\usepackage{amsmath}
\usepackage{latexsym}
\usepackage{longtable}
\usepackage{verbatim}
\usepackage{mathtools}
\usepackage{multirow}
\usepackage{empheq}
\usepackage{color}
\usepackage{longtable,booktabs}
\usepackage[none]{hyphenat}
\usepackage{setspace}
\usepackage{lineno}
\usepackage{graphicx}
\usepackage{txfonts}
\usepackage[figuresright]{rotating}
\usepackage[colorlinks = true,
            linkcolor = blue,
            urlcolor  = blue,
            citecolor = blue,
            anchorcolor = blue]{hyperref}

\begin{document}

   \title{Radiation mechanism of twin kilohertz quasi-periodic oscillations in neutron star low mass X-ray binaries}

   \titlerunning{On the radiation mechanism of twin kHz QPOs in NS-LMXBs}

   \author{Chang-Sheng Shi\thanks{E-mail: shics@hainanu.edu.cn}
          \inst{1}
          \and
           Guo-Bao Zhang
          \inst{2}
          \and
          Shuang-Nan Zhang
          \inst{3}
          \and
          Xiang-Dong Li
          \inst{4}
          }

   \institute{School of Physics and Optoelectronic Engineering, Hainan University, Hainan 570228, China
   \and
   Yunnan Observatory of Chinese Academy of Sciences, Kunming, 650216, China
   \and
  Key Laboratory of Particle Astrophysics, Institute of High Energy Physics, Chinese Academy of Sciences, Beijing 100049, China
     \and
  School of Astronomy and and Space Science, Nanjing University, Nanjing 210023, China\\}

    \date{}
   \date{Received ..., 202x; accepted ..., 202x}

  \abstract
   {The connection between quasi-periodic oscillations (QPOs) and magnetic fields has been investigated across various celestial bodies. Magnetohydrodynamics (MHD) waves have been employed to explain the simultaneous upper and lower kilohertz (kHz) QPOs.
   Nevertheless, the intricate and undefined formation pathways of twin kHz QPOs present a compelling avenue for exploration. This area of study holds great interest as it provides an opportunity to derive crucial parameters related to compact stars.
   }
   {This study strives to develop a self-consistent model elucidating the radiation mechanism of twin kHz QPOs, subsequently comparing it with observations.
  }
   {A sample of 28 twin kHz QPOs observed from the X-ray binary 4U 1636--53 are used to compare with the results of the MCMC calculations according to our model of the radiation mechanism of twin kHz QPOs, which is related to twin MHD waves.}
   {We obtain twenty-eight groups of parameters of 4U 1636--53 and a tight exponential fit between the flux and the temperature of seed photons to Compton up-scattering and find that the electron temperature in the corona around the neutron star decreases with the increasing temperature of the seed photons.}
   {The origin of twin kHz QPOs can be attributed to dual disturbances arising from twin MHD waves generated at the innermost radius of an accretion disc.
   The seed photons can be transported through a high temperature corona and Compton up-scattered. The variability of the photons with the frequencies of twin MHD waves can lead to the observed twin kHz QPOs.}


   \maketitle
%

\section{Introduction}
\noindent Quasi-Periodic Oscillations (QPOs) have been identified in many kinds of celestial objects including the Sun, stars, magnetars, neutron star low-mass X-ray binaries (NS-LMXBs), black hole LMXBs and active galactic nuclei \citep{van2006,Nakariakov2009,Shi2018}. However, the fastest X-ray variability, known as kilohertz QPOs (kHz QPOs), in NS-LMXBs have been paid more attention because the high frequencies of QPOs might be related to some effects of general relativity. KHz QPOs often emerge in pairs in the power spectra of NS-LMXBs, which are named as twin kHz QPOs including the upper and lower kHz QPOs according to their frequencies ($f_{\rm u}$ and $f_{\rm l}$).

Initially, the difference in the frequencies between the twin kHz QPOs appeared relatively constant, leading to the proposal of the beat-frequency model \citep{Miller1998}. In that model and subsequent models, it needs to be explained why the difference of twin kHz QPOs' frequencies is not a stochastic value and why they can emerge in the same time. Furthermore, \citet{Barret2005} and \citet{Ribeiro2017} posited that the origin of the upper QPOs was likely different from the lower QPOs due to the difference in some properties, including the fraction root-mean-square amplitude (hereafter rms), quality factors, time-lag between soft and hard X-ray photons. Therefore, a comprehensive explanation should account for both the relationship between these phenomena and their distinguishing characteristics.

 In LMXBs, the available models of QPOs can be mainly divided into two categories: dynamic mechanism and radiation origin. Generally, QPOs are supposed to be produced in an accretion disc or the surface of a corona, and numerous dynamic models have been proposed to account for their occurrence \citep[e.g.][]{Miller1998,Stella1999,Osherovich1999,Abramowicz2003,Shi2009,Shi2010,Shi20102,Ingram2011,Shi2014,Shi2018}.

In the context of radiation-based mechanisms, several key questions revolve around the specific locations where QPOs are generated and how they are ultimately produced in X-ray emissions. This involves understanding the connection between the aforementioned dynamic models and the actual generation of X-rays.
 \citet{Berger1996} obtained a correlation between the rms of QPOs and the energy of photons in 4U 1608$-$52. \citet{Zhang1996} discovered 870 Hz QPOs with the high values of the rms at high energies and suggested that the QPOs might originate either at the NS surface or very close to the NS because the temperature of the emission region of QPOs was higher than that of the accretion disc. \citet{Gilfanov2003} analysed the variability of  GX 340+0 and 4U 1608$-$52 and found that the rms spectrum of the QPOs was consistent with the Comptonized blackbody (BB) emission, which was assumed to originate from the boundary layer and not from the accretion disc.  \citet{Titarchuk2005} discovered that the temperature of the spectral component from a NS was higher than that from the disc in LMXBs. Thus a corona is often considered as the place where both the hard X-ray photons and QPOs originate.

\citet{Zhang2017} confirmed that the lower kHz QPOs appeared only during the transition from the hard state to the soft state, whereas the upper kHz QPOs emerged in most of the states in NS-LMXBs. It indicates that QPOs may depend on the states of the system and the rms of QPOs changes in different states. \citet{Lee2001} suggested that the resonance between the sources of soft photons and the Comptonizing medium might lead to the lower QPOs, which was fitted to the rms and the phase lag data of the kHz QPO with a frequency ($830\ \rm {Hz}$).
The increasing rms of QPOs with the increasing photon energy and a possible saturation or decline were observed in several works \citep{Berger1996, Yu1997, Gilfanov2003, Mukherjee2012}. The relation between the rms of QPOs and Comptonizing process were also discussed \citep[e.g. ][]{Lee1998, Zdziarski2005}. Then the Kompaneets equation was calculated in order to understand the energy dependence of rms and time lag of the lower kHz QPOs \citep{Kumar2014, Karpouzas2020}. However, a fundamental question that remains unanswered is how the simultaneous twin kHz QPOs are coupled into the X-ray radiation.

The variability in the Sun similar to QPOs are often called quasi-periodic pulsations (QPPs) and are often seen in different wave bands from radio to hard X-ray. QPPs have been attributed to potential origins in magnetohydrodynamic (MHD) oscillations, which may arise from standing modes within plasma structures or wave dispersion \citep{Nakariakov2009}. \citet{Nakariakov2004} suggested that fast and slow magnetoacoustic modes of a coronal loop might be the origin of QPPs in the Sun. The mechanism of QPOs in a NS-LMXB may be similar to that in the Sun, that is to say, QPOs may also be related to magnetohydrodynamics. However, it's essential to note that significant disparities exist in the gravitational forces and magnetic fields between the Sun and compact stars, thus the different values of gravity and magnetic field may lead to difference in the quasi-periods for QPPs in the Sun and QPOs in XBs.

\citet{Mendez2006} suggested that the frequencies of kHz QPOs might be determined at the accretion disc and their features might be modulated at the high-energy spectral component (e.g. corona/boundary layer).
A series of our previous works have been done on the relation between the frequencies of kHz QPOs and the accretion disc in LMXBs.
 \citet{Shi2009, Shi2010} first considered that the MHD waves were the origin of high frequency QPOs in NS-LMXBs and BH-LMXBs. Then the relation between an accretion rate and the frequencies of kHz QPOs were obtained by twin MHD waves \citet{Shi2014, Shi2018}. In order to explain the rms of twin kHz QPOs and combine the dynamic mechanism with the radiation process, we need to explore the physical process how twin MHD waves are produced at the innermost radius of an accretion disc and transported into the corona around a NS along with magnetic field lines \citep{Shi2009, Shi2014, Shi2018}.

In order to answer the above fundamental question, we combine the connection and the difference between upper and lower kHz QPOs, that is to say, we suppose that QPOs can be produced in the same dynamic mechanism but the twin QPO signals originate from different regions of the corona around the NS in the XB. The MHD waves interact with the plasma in the corona and lead to some oscillations (e.g. the heating rates, the temperature of electrons, the number density of photons) based on quasi-steady X-ray emission described by Kompaneets equation. Finally, the oscillations of the number density of emission photons lead to the X-ray variabilities in a power spectrum.
In this work, the basic physics to generate twin kHz QPOs in a corona and the basic equations to describe twin kHz QPOs
including Kompaneets equation are introduced in Sect. 2. In Sect. 3, the observational data from \citet{Zhang2017} are revisited in order to compare with our calculation by Monte Carlo method. The rms of 28 groups of twin kHz QPOs in 4U 1636$-$53 are also obtained.
Our discussion and conclusions about twin kHz QPOs are in the last section.\\

\section{Twin kHz QPOs produced in a corona} \label{sec:S2}
Studies by \citet{Shi2009, Shi2010} and \citet{Shi2014, Shi2018} have unveiled the presence of two distinct modes in magnetohydrodynamic (MHD) waves, offering an explanation for the concurrent upper and lower kHz QPOs.
Similar to the twin kHz QPOs, the twin MHD waves are named as the upper and the lower MHD wave according to their frequencies. Originating within the innermost regions of the accretion disc in a LMXB, these MHD waves propagate into the corona enveloping the NS within the LMXB (see Fig. 1).
As shown in \citet{Somov2012}, the upper MHD waves with higher frequencies can be transported in a shorter distance than the lower MHD waves (e.g. for slightly damped MHD waves, Landau damping).
To comprehend the intricate process underlying the generation of variability, we develop a radiation model wherein the longer wavelength mode permeates the NS surface entirely, while the shorter wavelength mode solely perturbs the outer layers, that is to say, the upper kHz QPOs emanate from the outermost layer of the corona, whereas the lower QPOs arise from the entire corona.
Finally, the radiation from Compton up-scattering will be changed in the form of a quasi-periodic variability due to oscillations from the MHD waves.
Consequently, the QPOs are posited to originate from the oscillating physical parameters of the plasma throughout the entire corona or from a specific segment of the corona influenced by disturbances from various physical processes associated with X-ray radiation in a quasi-steady state, which is the same to the steady radiation in the work of \citet{Kumar2014} and \citet{Karpouzas2020}.

A quasi-steady state is introduced here because maintaining an absolute steady state is challenging. The mass accretion rate's variability can affect the properties of the corona's surface region. However, the quasi-steady state of the radiation field expands to most of the corona and it is similar to the quasi-statics in the thermodynamics. This quasi-steady state can be maintained because the accumulation of accreting matter takes time and the macro parameters of the accretion system will hardly be changed in a time longer than milliseconds (the time scale of kHz QPOs). To align with the terminology used by \citet{Kumar2014} and \citet{Karpouzas2020}, we will refer to this quasi-steady state simply as the steady state below.

In principle, these perturbations can be generated by other mechanism, but the twin MHD waves are an attractive candidate as we have already shown that they generate roughly the correct frequencies \citep{Shi2009, Shi2014, Shi2018}.
In addition, the twin MHD waves offer an explanatory framework for why the two modes of twin kHz QPOs can be produced at different parts of the corona.

\begin{figure}[ht]
\begin{center}
\includegraphics[width=0.65\columnwidth]{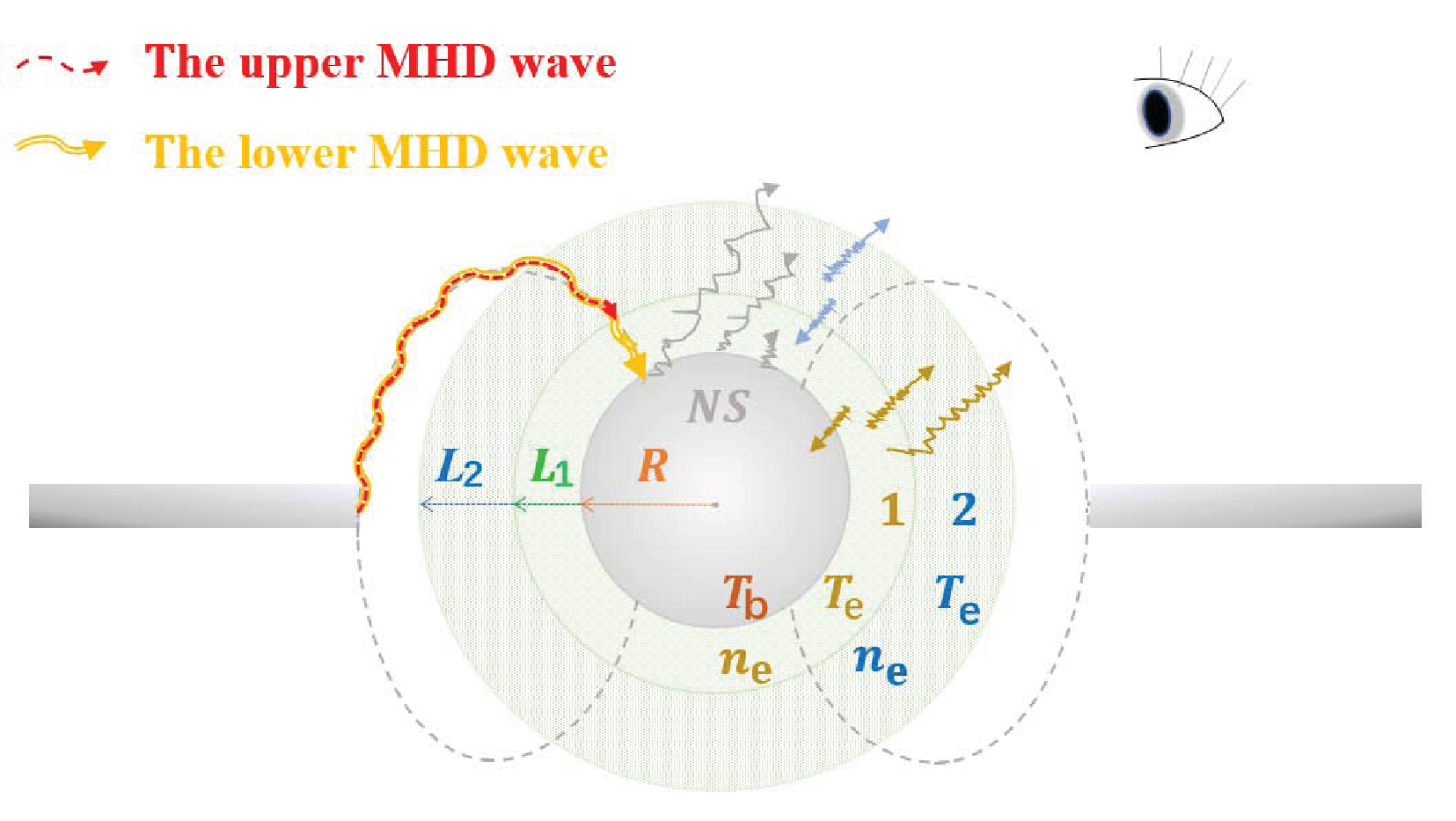}
  \caption{The sketch of the process on generating twin MHD waves at the innermost radius of the accretion disc in a LMXB, transporting the waves along the magnetic field lines, radiating X-ray photons from the surface of the corona. There are the same temperature and density in the whole corona in the steady radiation. The magnetic field lines are marked by the black dotted curves.}
  \label{fig1}
 \end{center}
\end{figure}

 Here we suppose that the canonical NS in a LMXB (the NS mass $M=1.4\ M_\odot$ and the NS radius $R\ =\ 10\ {\rm km}$) is around an uniform and isotropic spherical corona. Then we explore this model by perturbed Kompaneets equation for two layers of a corona below.

\subsection{Compton up-scattering in the two layers of a corona}

Suppose that the uniform spherical Comptonization corona with the depth $L$, electron temperature $T_{\rm e}$, and electron density $n_{\rm e}$ are around a cold NS as a spherical blackbody, which can release the seed photons and the injecting rate in per unit volume is,
\begin{equation}
\label{eq1}
{\dot{n}}_{s\gamma}=\frac{3R^2}{{\left(R+L\right)}^3-R^3}\left(\frac{2\pi{}}{h^3c^2}\frac{E^2}{e^{\frac{E}{kT_{b}}}-1}\right),
\end{equation}
where $T_{\rm b}$ is the temperature of BB radiation and the expression is appropriate for the whole corona \citep{Kumar2014,Karpouzas2020}.

Considering the upper MHD waves are transported only in the outmost layer near the surface of the corona, we divide the corona into two layers with the depths $L_1,\ L_2,$ and the volumes $V_1,\ V_2$ respectively, where the subscriptions $1$ and $2$ represent the first and the second layers respectively through this work  (Fig. 1). Then the whole optical depth of the corona can be expressed as $\tau_{{\rm T}}=\sum\limits_{i=1}^2 {n_{\rm e} L_i \sigma_{\rm T}}$, where $\sigma_{\rm T}$ is Thomson scattering cross section.

For the Comptonization that the low energy photons ($E \ll m_{\rm e}c^2$) are scattered in low energy electrons ($k T_{\rm e} \ll m_{\rm e}c^2$), it can be described by Kompaneets equation \citep{Kompaneets1957} and the updated form presented by \citet{Psaltis1997} , which is expressed as follows,

\begin{equation}
\label{eq2}
\begin{aligned}
& t_c\frac{\partial n_{\gamma{}}}{\partial t}=\frac{1}{{m_{\rm e}}c^2}\frac{\partial}{\partial E}\left[kT_{\rm e}\frac{\partial}{\partial E}\left(E^2n_{\gamma{}}\right)-4kT_{\rm e}En_{\gamma{}}+E^2n_{\gamma{}}\right]\\
& +t_c{\dot{n}}_{{\rm s}\gamma{}}-t_c\frac{c}{L}\frac{n_{\gamma{}}}{\left(1+\frac{1}{3}{\tau{}}_{\rm KN}\varepsilon\right)},
\end{aligned}
\end{equation}
where $m_{\rm e}$ is the electron mass, $c$ the light speed, $k$ the Boltzmann constant, $E$ the energy of photons, $t$ is time, $t_c=L/c\tau_{\rm T}$ Thomson collision timescale, $n_{\gamma}$ the photon number density per unit energy interval, $\tau_{\rm KN}$ (i.e. $\frac{\sigma_{\rm KN}}{\sigma_{\rm T}}\tau_{\rm T}$) the optical depth for Klein-Nishina cross section $\sigma_{\rm KN}$ and the parameter $\varepsilon$ depends on $E$ (see Appendix A). For simplicity, $\tau_{\rm KN}$ will be marked as $\tau$ and $\sigma_{\rm KN}$ will be marked as $\sigma$ below. Then $\tau,\ \tau_1,\ \tau_2$ are the whole optical depth, the local depth in the first and second layers with a Klein-Nishina cross section $\sigma,\ \sigma_1,\ \sigma_2$ respectively.

Similar to \citet{Karpouzas2020}, the probability per unit time for a photon escaping from the plasma with the depth ($L$) can be expressed as $P=\frac{c}{L(1+\frac{1}{3}{\tau{}}\varepsilon)}$ and the one corresponding to the depth ($L_i$) is $P_i =\frac{c}{L_i (1+\frac{1}{3}\tau_{i}\varepsilon)}$. Thus, the number of photons escaping from an unit volume per unit time can be described by the expression $P{n_{\gamma}}$ (or $P_i{n_{\gamma i}}$). In addition, the reciprocal of the probability means the average time for a photon to escape from the medium.

\subsection{Equations on the steady radiation and on the variabilities produced by MHD waves}
The twin kHz QPOs are produced by the oscillations from the disturbances of the MHD waves superposed on a steady X-ray radiation. After the MHD waves are transported into the corona, they will change the heating rates because the MHD waves can heat the plasma in the corona and thus produce oscillations. The change of the heating rate will change the electron temperature and other parameters. Then those disturbed parameters will change the final X-ray radiation and twin kHz QPOs will appear. Now the result after the disturbances will be introduced in the steady state below.

\subsubsection{The lower kHz QPOs originating from the disturbance in a whole corona}
The lower kHz QPOs are supposed to be generated from a disturbance over the whole corona in a steady state. A steady radiation is not related with time, that is, $\frac{dn_{\gamma{}0}}{dt}=0$, where the subscription 0 denotes the steady state. Thus the Kompaneets equation (Eq. 2) in the steady state in the whole corona can be simplified as,
 \begin{equation}
 \label{eq3}
 \begin{aligned}
&\frac{{\partial}^2n_{\gamma{}0}}{d E^2}+\frac{1}{kT_{\rm e0}}\frac{d n_{\gamma{}0}}{d E}-{\frac{2}{E^2}}n_{\gamma{}0}
+\frac{2}{EkT_{\rm e0}}n_{\gamma{}0}-t_{\rm c}\frac{{m_{\rm e}}c^2}{E^2kT_{\rm e0}}\frac{c}{L}\frac{n_{\gamma{}0}}{(1+\frac{1}{3}{\tau{}}_{\rm 0}\varepsilon)}\\
&=-\frac{1}{{l E^2}}t_{\rm c}{\dot{n}}_{\rm s\gamma{}0},
\end{aligned}
\end{equation}
where $l=\frac{kT_{\rm e0}}{{m_{\rm e}c^2}}$ and Eqs. 3-6 can be obtained from \citet{Karpouzas2020}.

For an electron system, the change rate of its temperature should be attributed to the competition between the heating rate and the cooling rate, which can be expressed as,
 \begin{equation}
\label{eq4}
\frac{\partial{}(\frac{3}{2}kT_{\rm e0})}{\partial{}t}={\dot{H}}_{\rm ex0}-<\Delta \dot{E}>,
\end{equation}
where ${\dot{H}}_{\rm {ex}}$ is  the external heating rate per electron, and $<\Delta \dot{E}>$ is the Compton cooling rate per electron, which can be expressed as $\int_{E_{\rm min}}^{E_{\rm max}}\frac{n_{\gamma{}0}{\sigma{}}_{\rm T}}{m_{\rm e}c}(4kT_{\rm e}-E)EdE$ and the limits (${E_{\rm min}},\ {E_{\rm max}}$) are assumed as $2\ {\rm keV}$ and $60\ {\rm keV}$, which are the same with \citet{Kumar2014} and \citet{Karpouzas2020}. In the steady state, the temperature of the electron system will be kept unchanged ($\frac{\partial{}T_{\rm e0}}{\partial{}t}=0$), then the heating rate can be obtained as,
 \begin{equation}
  \label{eq5}
{\dot{H}}_{\rm ex0}=\frac{1}{m_{\rm e}c}\int_{E_{\rm min}}^{E_{\rm max}}{\sigma{}}(4kT_{\rm e0}-E)n_{\gamma{}0}EdE.
\end{equation}

Considering that some photons impinge back onto the seed photons sources with the feedback coefficient ($\eta$) , the energy of those photons is $\eta{}V \int P n_{\gamma{}0}EdE$. Then the steady radiation in the border between the NS and the whole corona can be obtained as
 \begin{equation}
\label{eq6}
4\pi{}R^2{\sigma{}}_{\rm T}T_{\rm b0}^4-\eta{}V \int P n_{\gamma{}0}EdE=4\pi{}R^2{\sigma{}}_{\rm T}{T_{\rm b0}^{'}}^4,
\end{equation}
where $4\pi{}R^2{\sigma{}}_{\rm T}{T_{\rm b0}^{'}}^4$ is the energy generation rate inside the seed photon source and it is nearly unchanged \citep{Kumar2014,Karpouzas2020}.

After the disturbance from the lower MHD wave, some variables can be expressed as,
$T_{\rm el}=T_{\rm e0}(1+\Delta{}T_{\rm e0}e^{-i{\omega{}}_1t})$,
$T_{\rm bl}=T_{\rm b0}(1+\Delta{}T_{\rm b0}e^{-i{\omega{}}_1t})$,
$n_{\gamma l}=n_{\gamma 0}(1+\Delta{}n_{\gamma 0}e^{-i{\omega{}}_1t})$,
$\dot{H}_{\rm exl}=\dot{H}_{\rm ex0}(1+\Delta{}\dot{H}_{\rm exl}e^{-i{\omega{}}_1t})$, where $\omega{}_1=2\pi f_{\rm l}$, $T_{\rm el},\ n_{\gamma {\rm l}},\ T_{\rm bl},\  \dot{H}_{\rm exl} $ are electron temperature, electron density, the temperature of seed photons and heating rate after the disturbance.
Combing equations (3), (4), (6), the equations on the disturbance can be obtained as,
 \begin{equation}
  \label{eq7}
\begin{array}{lll}
-\frac{\partial^2\Delta{}n_{\gamma{}0}}{\partial E^2}-\frac{1}{kT_{e0}}\frac{\partial \Delta{}n_{\gamma{}0}}{\partial E}
-\frac{2}{n_{\gamma{}0}}\frac{\partial \Delta{}n_{\gamma{}0}}{\partial E}\frac{\partial n_{\gamma{}0}}{\partial E}
+t_{\rm c}\frac{1}{{}n_{\gamma{}0}}\frac{{\dot{n}}_{{\rm s}\gamma 0}}{l E^2}\Delta{}n_{\gamma{}0}\\
+\frac{1}{l E^2}(-i\omega_1)t_{\rm c}\Delta{}n_{\gamma{}0}\\
 =\frac{-2}{E^2}\Delta{}T_{\rm e0}+\frac{1}{n_{\gamma{}0}}\frac{d^2 n_{\gamma{}0}}{dE^2}\Delta{}T_{\rm e0}+
\frac{t_{\rm c} {\dot{n}}_{s\gamma 0}}{l E^2 n_{\gamma{}0}}\frac{{E}/{kT_{\rm b0}}}{1-exp(-{E}/{kT_{\rm b0}})}\Delta{}T_{\rm b0},
\end{array}
\end{equation}

 \begin{equation}
  \label{eq8}
\Delta{}T_{\rm b0}=\frac{{\eta{}V}_{c1}}{16\pi{}a^2{\sigma{}}_TT_{\rm b0}^4}
\int\frac{n_{\gamma{}0}{\Delta{}n_{\gamma{}0}}}{t_c{\tau{}}_T(1+\frac{1}{3}{\tau{}}_{\rm 0}\varepsilon)}EdE,
\end{equation}

 \begin{equation}
  \label{eq9}
  \begin{aligned}
&\Delta{}T_{\rm e0}=({\dot{H}}_{\rm ex0}{\Delta{}{\dot{H}}_{exl}}-\frac{ kT_{\rm e0}}{m_{\rm e}c}
\int_{E_{\rm min}}^{E_{\rm max}}{\sigma{}}{4n_{\gamma{}0}\Delta{}n_{\gamma{}0}}EdE\\
&+\frac{1}{m_{\rm e}c}\int_{E_{\rm min}}^{E_{\rm max}}{{\sigma{}}_{\rm }n}_{\gamma{}0}{\Delta{}n_{\gamma{}0}}E^2dE)\\
&/[\left(-i{\omega{}}_1\right){\frac{3}{2}}kT_{e0}+4\frac{kT_{\rm e0}}{m_{\rm e}c}\int_{E_{\rm min}}^{E_{\rm max}}
{\sigma{}}{n_{\gamma{}0}}EdE].
\end{aligned}
\end{equation}

\subsubsection{The upper kHz QPOs originating from the perturbation in the outermost layer of a corona}
The upper MHD waves will only interact with the plasma in the second layer of the corona after the waves are transported into the corona, thus the corona are considered to be divided into two layers.
Due to that the steady state for the corona is stationary, the seed photon number inputting the corona should be same regardless if the corona is layered or not.
In subsection 2.2.1, seed photons are injected into the medium with a rate ($\dot{n}_{{\rm s}\gamma}$), which is averaged in the whole corona in the same way done previously \citep{Kumar2014,Karpouzas2020}. After the corona is layered, the seed photons will injected into the first layer with the injection rate ${\dot{n}}_{\rm s\gamma{}1,0}={\dot{n}}_{\rm s\gamma{}0}\frac{V}{V_{1}}$.

Similar to equation (3), the Kompaneets equation in the two layers in the steady state can be simplified as,
 \begin{equation}
 \label{eq10}
  \begin{aligned}
& \frac{{\partial{}}^2n_{\gamma{}1,0}}{\partial{} E^2}+\frac{1}{kT_{\rm e0}}\frac{\partial{} n_{\gamma{}1,0}}{\partial{} E}-{\frac{2}{E^2}}n_{\gamma{}1,0}
+\frac{2}{E kT_{\rm e0}}n_{\gamma{}1,0}\\
&-t_{\rm c}\frac{1}{l E^2}\frac{c}{L_1}\frac{n_{\gamma{}1,0}}{(1+\frac{1}{3}{\tau{}}_{\rm 1}\varepsilon)}
=-\frac{1}{{l E^2}}t_{\rm c}{\dot{n}}_{\rm s\gamma{}1,0},
\end{aligned}
\end{equation}
and
 \begin{equation}
 \label{eq11}
  \begin{aligned}
& \frac{{\partial{}}^2n_{\gamma{}2,0}}{\partial{}E^2}+\frac{1}{kT_{\rm e0}}\frac{\partial{}n_{\gamma{}2,0}}{\partial{}E}-{\frac{2}{E^2}}n_{\gamma{}2,0}
+\frac{2}{EkT_{\rm e0}}n_{\gamma{}2,0}\\
&-t_{\rm c}\frac{1}{l E^2}\frac{c}{L_2}\frac{n_{\gamma{}2,0}}{(1+\frac{1}{3}{\tau{}}_{\rm 2}\varepsilon)}
=-\frac{1}{{l E^2}}t_{\rm c}{\dot{n}}_{\rm s\gamma{}2,0},
\end{aligned}
\end{equation}
where ${\dot{n}}_{\rm s\gamma{}2,0}=\frac{c}{L_1}\frac{n_{\gamma{}1,0}}{(1+\frac{1}{3}{\tau{}}_{\rm 1}\varepsilon)}\frac{V_1}{V_{2}}$  is the density of the input seed photons in the second layer per unit time, which comes from the output photons in the first layer. The energy changing rate of per electron in each layer of the corona can be expressed as,
 \begin{equation}
\label{eq12}
\frac{\partial{}(\frac{3}{2}kT_{{\rm e}i, 0})}{\partial{}t}={\dot{H}}_{{\rm ex}i,0}-\int_{E_{\rm min}}^{E_{\rm max}}\frac{n_{\gamma{}i,0}{\sigma{}}_{\rm T}}{m_{\rm e}c}(4kT_{{\rm e} 0}-E)EdE.
\end{equation}


Similar to equation (5), the energy that is released and collected in each layer is also balanced, that is, $\frac{\partial{}T_{{\rm e}i,0}}{\partial{}t}=0$, thus the heating rate in the two layers is expressed as,
 \begin{equation}
  \label{eq13}
{\dot{H}}_{{\rm ex} i,0}=\frac{1}{m_{\rm e}c}\int_{E_{\rm min}}^{E_{\rm max}}{\sigma{}}_{i}(4kT_{\rm e0}-E)n_{\gamma{}i,0}EdE.
\end{equation}

Because the whole corona and every part are in steady state, radiation from every boundary should be kept unchanged and the equation from the steady radiation in the boundary between the two layers is,
 \begin{equation}
\label{eq14}
V_1 \int P_1 n_{\gamma{}1,0}EdE-\eta{}V_2 \int P_2 n_{\gamma{}2,0}EdE=Const,
\end{equation}
where $V_1 \int P_1 n_{\gamma{}1,0}EdE$ is the outward radiation energy from the first layer and $\eta{}V_2 \int P_2 n_{\gamma{}2,0}EdE$ the photon energy bounced back from the second layer.

After the disturbance, the variables can be expressed as,
$T_{\rm eu}=T_{\rm e0}(1+\Delta{}T_{\rm e u}e^{-i{\omega{}}_2t})$,
$n_{\gamma \rm u}=n_{\gamma 2,0}(1+\Delta{}n_{\gamma 2,0}e^{-i{\omega{}}_2t})$,
$\dot{H}_{\rm exu}=\dot{H}_{\rm ex2,0}(1+\Delta{}\dot{H}_{\rm exu}e^{-i{\omega{}}_2t})$, where $\omega{}_2=2\pi f_{\rm u}$. Similar to the change of $T_{\rm b}$ from ${\dot{n}}_{\rm s\gamma{}0}$ after being disturbed by the lower wave, ${\dot{n}}_{\rm s\gamma{}2,0}$ should also be changed by the upper wave. The change is from the disturbance of the effective density of seed photons in the second layer $n_{\gamma{}{\rm us}}=n_{\gamma{}1,0}(1+\Delta {n_{\gamma{}{\rm us}}e^{-i\omega_2t})}$, which shows the interaction of the upper wave with the seed photons entering the second layer (${\dot{n}}_{\rm s\gamma{}2,0}$).

Then the change of photon density in the second layer can be obtained from equation (11) and expressed as,
 \begin{equation}
  \label{eq15}
\begin{aligned}
&-\frac{\partial^2\Delta{}n_{\gamma{}2,0}}{\partial E^2}-\frac{1}{kT_{e0}}\frac{\partial \Delta{}n_{\gamma{}2,0}}{\partial E}
-\frac{2}{n_{\gamma{}2,0}}\frac{\partial \Delta{}n_{\gamma{}2,0}}{\partial E}\frac{\partial n_{\gamma{}2,0}}{\partial E}\\
&+t_c\frac{{\dot{n}}_{{\rm s}\gamma 2,0}}{{l E^2 n}_{\gamma{}20}}\Delta{}n_{\gamma{}2,0}+t_c\frac{1}{l E^2}
(-i\omega_2)\Delta{}n_{\gamma{}2,0}\\
& =\frac{-2}{E^2}\Delta{}T_{\rm eu}+\frac{1}{n_{\gamma{}2,0}}\frac{\partial^2n_{\gamma{}2,0}}{\partial E^2}\Delta{}T_{\rm eu}+
\frac{t_{\rm c}}{l E^2 n_{\gamma{}2,0}}\frac{V_1}{V_2}\frac{c}{L_1}\frac{n_{\gamma{}1,0}}{(1+\frac{1}{3}{\tau{}}_{\rm 1}\varepsilon)}\Delta{}n_{\gamma{}{\rm us}}.
\end{aligned}
\end{equation}

According to equations (12) and (14), the changes of the electron temperature and that of the density of seed photons in the second layer are,
 \begin{equation}
  \label{eq16}
  \begin{aligned}
&\Delta{}T_{eu}=({\dot{H}}_{ex2,0}{\Delta{}{\dot{H}}_{exu}}-4lc{\sigma{}}_T\int_{Emin}^{Emax}n_{\gamma{}2,0}{\Delta{}n_{\gamma{}2,0}}EdE\\
&+\frac{{\sigma{}}_{\rm T}}{m_ec}\int_{Emin}^{Emax}n_{\gamma{}2,0}{\Delta{}n_{\gamma{}2,0}}E^2dE)\\
&/[\frac{3}{2}
kT_{e0}\left(-i{\omega{}}_2\right)+4lc{\sigma{}}_T\int_{Emin}^{Emax}n_{\gamma{}2,0}EdE],
\end{aligned}
\end{equation}
and
 \begin{equation}
  \label{eq17}
{\Delta{}n}_{\gamma{}\rm us}={\eta{}} V_{2}\int P_2 n_{\gamma{}2,0}\Delta{}n_{\gamma{}2,0}EdE/\int P_1 V_{1}n_{\gamma{}1,0}EdE.
\end{equation}

Combining equations (3), (10), (11), the density of the diffusion photons ($n_{\gamma{}0},\ n_{\gamma{}1,0},\ n_{\gamma{}2,0}$) can
be calculated. Then the disturbed density of photons ($\Delta{}n_{\gamma{}0}$) can also be obtained from equations (7), (8), (9)
and $\Delta{}n_{\gamma{}2,0}$ from equations (15), (16), (17). The parameters $n_{\gamma{}0},\ n_{\gamma{}1,0},\ n_{\gamma{}2,0}$, $\Delta{}n_{\gamma{}0},\ \Delta{}n_{\gamma{}2,0}$
determine the rms of twin kHz QPOs, which can be expressed as ${\rm RMSL}=\int  n_{\gamma{}0}\Delta{}n_{\gamma{}0}dE/\int n_{\gamma{}0}dE$
for the lower QPOs and ${\rm RMSU}=\int  n_{\gamma{}2,0}\Delta{}n_{\gamma{}2,0}dE/\int n_{\gamma{}2,0}dE$ for the upper QPOs \citep[also see][]{Kumar2014,Karpouzas2020}.
The expressions of rms of twin kHz QPOs clearly show that QPOs depend on the states of sources, which can be used to study the
properties of sources.

\section{The key physical variables obtained by Monte Carlo method} \label{sec:S4}
\subsection{The basic parameters to match the observation}
In order to compare with observations, we revisit the observational data of the twin kHz QPOs in 4U 1636$-$53 obtained by \citet{Zhang2017}, in which the energy spectra in the energy interval ($2\sim60\ {\rm keV}$) are used to determine the state parameters of the LMXB.
The calculated energy spectrum with the expression $f(E)\ =\ \frac{c}{L}\frac{n_{\gamma{}0}}{\left(1+\frac{1}{3}{\tau{}}\varepsilon\right)}*V/4\pi r^2$ (in units of Photons s$^{-1}$ keV$^{-1}$ cm$^{2}$) is compared to the observed spectra, where $r$ is the distance from 4U 1636$-$53 to the earth estimated as $6.0\pm0.5\ {\rm kpc}$ by \citet{Galloway2006}.
According to the energy spectrum, the flux can be obtained by the expression, $\int_2^{60}f(E)EdE$.
In addition, the fitted models for rms of kHz QPOs in \citet{Ribeiro2017}(${\rm RMSL}=8.32(\pm0.05)*e^{-\frac{[f_{\rm l}-755.9(\pm1.5)]^2}{2*[150.74(\pm 2.12)]^2}}$; ${\rm RMSU}=20(\pm 1)-0.0148(\pm0.0008) f_{\rm u}+3.2(\pm0.3)*e^{-\frac{[f_{\rm u}-799(\pm11)]^2}{2*[97(\pm13)]^2}}$) are used to obtain the rms of the 28 twin kHz QPOs.


\citet{Lin2007} summarized many models to fit the energy spectra of accreting NS-LMXBs and most models
included a soft thermal component and a hard Comptonized component. In our model, the NS's BB photons are
Comptonized and then most of the BB photons do not escape from corona, which means that the component of the NS's BB photons in the spectrum is weak.
In the work of \citet{Zhang2017}, fitting the spectrum of 4U 1636$-$53 revealed that, above 4 keV, both the accretion disc component and the directly observed BB component from the neutron star are much weaker compared to the Comptonized component.
Again in our model, the disc component is not Comptonized effectively due to the much smaller coverage of the corona to the disc. Therefore, it is reasonable to exclude the soft component in our fitting. However, this exclusion may cause our fitted results for the lower energy spectrum ($E  \lesssim 4\ \rm{keV}$ and around $6.4\ \sim\ 7\ \rm{keV}$) deviate from the observations, which will be discussed further in Sect. 4.2.
Then we have chosen to narrow our focus to specific observed parameters, including the frequencies, RMSL, RMSU, and flux density from the spectra.


\subsection{The five parameters obtained in a steady state and the three parameters in a disturbance}
In order to compute the rms of QPOs, it is necessary to provide eight essential parameters that need to be input to the equations in Sect. 2: five system parameters ($kT_{\rm e},\ kT_{\rm b},\ \tau,\ L,\ L_2$), one physical parameter ($\eta$) and two perturbation parameters ($\Delta{}\dot{H}_{\rm exl},\ \Delta{}\dot{H}_{\rm exu}$). Besides those parameters, the frequencies of the observed twin kHz QPOs ($f_{\rm l}$, $f_{\rm u}$) should also be input to the equations in Sect. 2.

The analysis by \citet{Zhang2017} indicates that the lower kHz QPOs across a much broader range of optical depths compared to the upper kHz QPOs. Additionally, the range of optical depth values for the upper kHz QPOs falls within the wider range observed for the lower kHz QPOs. In other words, most of the
optical depths associated with the lower QPOs are generally considerably higher than those linked to the upper QPOs.
Statistically, this implies that individual lower QPOs may be indicative of a thick corona, while individual upper QPOs may correspond to a thin corona, assuming a slight change in electron density ($n_{\rm e}$), that is to say, the plasma layer where the lower QPOs emerged is thicker than the one where upper QPOs emerged, which is consistent with the result of the transported depth of twin MHD waves in a corona.
Consequently, it is suggested that the lower QPOs are presumed to originate from perturbations throughout the entire corona, while the upper kHz QPOs arise from disturbances in the outermost layer of the corona. Thus, other parameters can be obtained from those eight parameters, including $\tau_2=L_2 \tau/L,\ L_1=L-L_2$ and $\tau_1=\tau-\tau_2$.


The parameters in the steady states ($kT_{\rm e},\ kT_{\rm b},\ \tau,\ L$) can be obtained by MCMC fitting based on Eqs. 3-6 according to
matching the spectra ($f(E)$). The steady states also can be described by a multi-layers corona approximately \citep{Shi2021}, thus those parameters and $L_2$ can also be obtained by MCMC fitting based on Eqs. 10-14.

The above five parameters can be fitted by the spectra in a steady state, but the sole values of $\Delta{}\dot{H}_{\rm exl},\ \Delta{}\dot{H}_{\rm exu},\ \eta$ in a disturbance can not be obtained due to the unknown feedback factors ($\eta$). Thus, the heating rate and a feedback factor ($\eta\in (0, 1)$) will be estimated to align with the observational data, and their characteristic values will then be calculated. First, the density of diffusion photons ($n_{\gamma{}0},\ n_{\gamma{}1,0},\ n_{\gamma{}2,0}$), the variables ($\Delta{}n_{\gamma{}0},\ \Delta{}n_{\gamma{}2,0}$) can be calculated with equations (7)-(9) and (15)-(17) when $\Delta{}\dot{H}_{\rm exl},\ \Delta{}\dot{H}_{\rm exu},\ \eta$ are chosen. Then RMSL and RMSU can be obtained according to all the parameters.
Finally RMSL and RMSU of one twin kHz QPOs can be used to match the observation and to determine the value of the heating rate for a special $\eta$.

In summary, the acquisition of the eight parameters can be achieved through the following procedures:
 \begin{enumerate}
\item The four parameters ($kT_{\rm e},\ kT_{\rm b},\ \tau,\ L$) are obtained by the MCMC method for the NS system in a steady state. Specifically, a model energy spectrum is first generated with Eqs. 3-6 with the four parameters sampled in the 4-dimensional parameter space, and then compared with the observed spectrum by minimizing $\chi^2$.
     We used over 50000 iterations for every one of the 28 observations with the twin kHz QPOs listed in Table B.1.
\item Similar to the above first fit, the depth of the second layer ($L_2$) is obtained by the MCMC method according to Eqs. 10-14 in the above steady state. We used over 10000 iterations after substituting the central values of $kT_{\rm e},\ kT_{\rm b},\ \tau,\ L$ from the above fitting. Of course, the above five parameters can also be obtained by the MCMC method. However, due to the extensive computational effort required for MCMC and the approximate nature of the results regarding the layering of a corona (as discussed in \citet{Shi2021}), we have chosen to focus on obtaining $L_2$ after substituting the four parameters.
\item Choose a feedback factor changed from 0 to 1 in a step length 0.1, substitute the central values of $kT_{\rm e},\ kT_{\rm b},\ \tau,\ L,\ L_2$ from the above steps 1 and 2, then choose 3000 random heating rate in $(0,30)$, calculate RMSL and RMSU and select the ones within the error limits of RMSL and RMSU.
\item Select the feedback factor for the largest number heating rate to make RMSL and RMSU within their error limits. Select the twin heating rate with the chosen $\eta$ by matching the central value of RMSL (or RMSU) as the characteristic values.
\end{enumerate}

In the method of calculation, equations (3), (5), (11), (13), (7)-(9), (15)-(17) need to be rewritten into linear difference equations with some mathematical techniques (second-order accurate central differences formulas, Simpson rule) used by \citet{Karpouzas2020}. The natural energy boundary conditions ($n_{\gamma},\ n_{\gamma i}=0$ for $E_{\rm min}$, $E_{\rm max}$) are also applied to the obtained difference equations.

Finally, the five parameters fitted in a steady state and the three characteristic parameters in a disturbance are listed in Table B.1, which has the similar values to \citet{Karpouzas2020}.
However, it should be noted that the method on layering the corona into the two layers is an approximate one, and the effective depth of the outmost layer ($L_2$) is the upper limit that the above method on layering the corona in a steady state can be used to calculate RMSU. The real penetration depth of the upper MHD wave is determined by the specific interaction between the waves and the plasma, which needs to be explored in future in MHD. Thus the effective depth ($L_2$) is used to calculate the characteristic values of $\Delta{}\dot{H}_{\rm exl},\ \Delta{}\dot{H}_{\rm exu},\ \eta$.

As described above, the change of the number density of photons from the heating rate can be obtained if the interaction between MHD waves and plasma is understood. With this information, the rms can be calculated, making it possible to reproduce the observed characteristics of the studied kHz QPOs. For example, the relationship between rms and optical depth ($\tau$), as depicted in Fig. 3 by \citet{Ribeiro2017}, can be modeled accurately.

\subsection{Our results of MCMC fitting}
\subsubsection{The temperature of seed photons and the four characteristic parameters}
Finally, the eight parameters for every group of twin kHz QPOs are obtained by fitting with the observed energy spectrum and then comparing with the observed RMSL and RMSU (see Table B.1). In those solutions, all the values of $k T_{\rm b}$ are in a small region (i.e $0.19-0.22\ \rm keV$), which is within the range obtained from the spectral fitting result \citep{Lin2007} and from the work of \citet{Karpouzas2020}. This suggests that it is reasonable to assume that the NS is encircled with a corona and the energy exchange happens mainly in the corona. However, a strong degeneracy between cold and hot-seed photon models is shown in \citet{Lin2007} and \citet{Karpouzas2020}. Although the seed photons in this study are cold, it remains to be explored whether these cold seed photons originate from the accretion disc and the hot seed photons from the NS when the accretion disc is introduced into our model.
A positive correlation between flux and $k T_{\rm b}$ for the 28 twin kHz QPOs is also shown in panel (a) of Fig. 2.
According to the fitting result, there is a tight exponential fit between the flux and the temperature of seed photons, that is to say, the flux can be expressed as, $5.97*10^{-13}*\rm{Exp}{(53.26*{kT_{\rm b,\ \rm kev}})}\ {\rm erg*s^{-1}*cm^{-2}}$.

In panel (b), the constant function ($2.96\pm0.04$) and the linear function (with the slope $-48.45\pm8.05$ and the intercept $-13.04\pm1.67$) are obtained by fitting the data when the errors are considered. According to the results, the linear function is statistically more suitable than the constant function for panel b, although both the two reduced chi-squares (11.78 and 27.17) from the fits are big.
Thus, a weak negative correlation between $k T_{\rm e}$ and $k T_{\rm b}$ can be found. In panel (c), rms for lower QPOs and upper QPOs decrease with the increasing $k T_{\rm b}$.
In panel (d), rms ratio (${\rm RMSL}/{\rm RMSU}$) increases slightly with the increasing $k T_{\rm b}$ but the trend is not obvious due to the large error of rms ratio. There are similar reduced chi-squares for the fits using the constant function ($\rm {rms\ ratio} = 0.58\pm 0.06$) and using the linear function ($\rm {rms\ ratio} = (25.01\pm 13.80)*kT_{\rm b}- (4.42 \pm 2.76)$), which are $0.98$ and $1.06$ respectively. The big errors of the fitted parameters originate from the data with big errors, which means that an obvious linear correlation between the rms ratio and $kT_{\rm b}$ is not discovered.

\begin{figure*}[!htbp]
\begin{center}
\includegraphics[width=1.8\columnwidth]{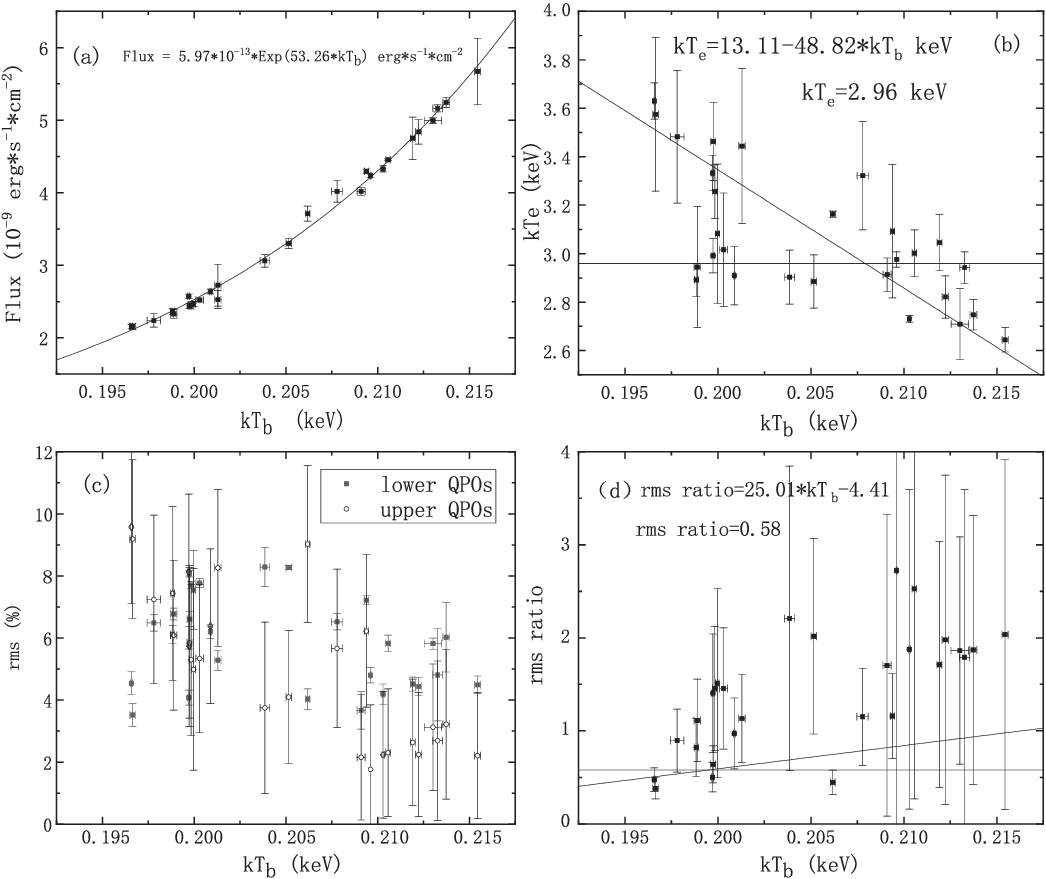}
  \caption{The relation between the temperature of the seed photons on the NS and several variables ( a: the flux of 4U 1636--53 with different twin kHz QPOs; b: the temperature of the electrons in the corona; c: the rms of twin kHz QPOs; d: rms ratio of lower QPOs to upper QPOs).}
  \label{fig5}
 \end{center}
\end{figure*}

\subsubsection{The effect of electron temperature on QPOs and states}
In panels (a) and (b) of Fig. 3, there are negative correlations between the frequencies (or frequency ratio) of QPOs and electron temperature. As known by us, direct relation between rms of QPOs and the electron temperature have not been discussed in theory; however, indirect dependencies between them as well as the state of the source may exist. It has been discussed that the frequencies of QPOs will go up with the decreasing innermost radius of an accretion disc \citep[e.g. ][]{Shi2014, Shi2018}. In addition, the innermost radius of an accretion disc is related to the state of the source. Thus the frequencies (or frequency ratio) of QPOs may be related to electron temperature indirectly. After carrying on linear fits, we find that the negative correlations between the frequencies of QPOs and electron temperature are weaker than that between frequency ratio and electron temperature. It means that the state of the source is more strongly associated with the frequency ratio than either the lower or upper QPO alone.

In panel (c) of Fig. 3,
there is a complex relation between the rms ratios of twin QPOs and the electron temperature, which may originate from the perturbed state. As discussed, rms is related $n_{{\gamma}0}$ and $\delta n_{{\gamma}0}$ which can be influenced by the system's response to external perturbations.
The state of the source leads to different distributions of spectra. Thus rms ratios will be different for the different distributions of spectra and is affected by the state, $T_{\rm e}$ and external perturbation.

\begin{figure*}[!htbp]
\begin{center}
\includegraphics[width=2.0\columnwidth]{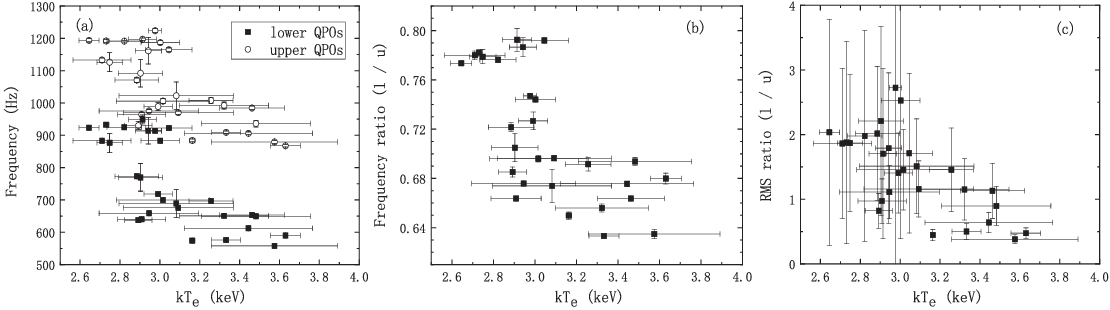}
  \caption{The relation between the temperature of the electrons in the corona and the variables ( a: frequencies of QPOs; b: the frequency ratio of the lower QPOs to that of the upper QPOs; c: rms ratio of lower QPOs to upper QPOs).}
  \label{fig8}
 \end{center}
\end{figure*}

\subsubsection{The ratio of the change in the heating rate}
The corona will be heated after twin MHD waves are transported into it and will be disturbed so as to form the twin kHz QPOs. As shown in panel (a) of Fig. 4, the ratio of the change in the heating rate due to the twin MHD waves (${\delta \dot{H}_{\rm L}}\ / \ {\delta \dot{H}_{\rm U}}$) increases with the ratio of the twin frequencies ($f_{\rm L}\ / \ f_{\rm U}$). It means that the frequency ratio will affect ${\delta \dot{H}_{\rm L}}\ / \ {\delta \dot{H}_{\rm U}}$. Essentially, the twin MHD waves will share the exchanged energy with the plasma. In addition, the frequency ratio is related to the propagation depth of MHD waves, which is affected by the damping of these waves due to energy dissipation mechanisms, such as Landau damping. However, the energy dissipation is also related to the density of plasma and a good relation between $f_{\rm L}\ / \ f_{\rm U}$ and $L\ / \ L_2$ is not found. Therefore, further investigation into the interaction between MHD waves and plasma is both intriguing and valuable.

In Panel (b) of Fig. 4, there is a negative correlation between $\eta$ and ${\delta H_{\rm L}}\ / \ {\delta H_{\rm U}}$. As the feedback coefficient increases, more energy is redirected back into the seed sources, enhancing the significance of energy exchange from the lower frequency waves. Consequently, the rate of change of the heating rate for the lower frequency waves becomes higher than that for the upper frequency waves. The possible reason is that the lower waves lose energy more easily than the upper waves and it needs to be explored.

\begin{figure*}[!htbp]
\begin{center}
\includegraphics[width=2.0\columnwidth]{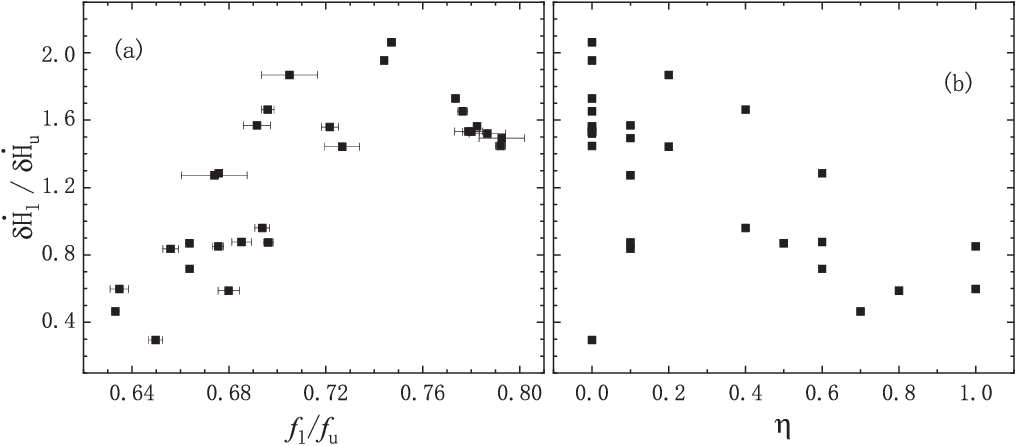}
  \caption{The relations between the ratio of the changing rate of heating rate (${\delta H_{\rm L}}\ / \ {\delta \ H_{\rm U}}$) and the ratio of QPO frequencies, feedback coefficients when the central values of RMSL and RMSU are considered.}
  \label{fig12}
 \end{center}
\end{figure*}

\subsection{Single parameter dependence}
According to the expression on rms of QPOs and flux, the parameters that affect $n_{\gamma}$ and $\Delta n_{\gamma}$ also have effect on the rms of QPOs and flux. However, some parameters to be used to calculate the rms of QPOs and flux should be coupled with another (e.g. $k T_{\rm b}$ and $k T_{\rm e}$, $f$ and $L$, $L$ and $\tau$), which lead to the complexity of QPOs' study and the incomplete consistency when only one parameter is changed below.

However, the results from changing one parameter also have some similarity and may find the key factor affecting the radiating system. Thus the flux and rms ratio of twin kHz QPOs are calculated below when only one of the eight parameter is changed, while keeping the others constant as listed in Table B.1 for the same twin QPOs. In order to compare to the observation, the flux is calculated between $4\sim23\ \rm{keV}$ in this subsection. The unabsorbed bolometric flux as the observed result in this study is obtained by using the model outlined in Zhang et al. (2017), accounting for the impact of interstellar absorption, with the column density fixed at $N_{\rm H} = 3.1 \times 10^{21}\ {\rm {cm}^{-2}} $. As an example, the central values of the eight parameters for the twin kHz QPOs with the frequencies ($\nu_{\rm l}=883.44\pm{1.10}\ {\rm Hz}, \nu_{\rm u}=1132.73\pm6.47\ {\rm Hz}$) in Table B.1 are used in this subsection except the data marked by "plus" and "cross" in Fig. 5.

As seen in panel (a) of Fig. 5, the consistency of the observed flux and the one from our model shows that the fitted results are reasonable. To understand dependence of flux on $k T_{\rm b}$ only, we vary $k T_{\rm b}$ while fixing all other parameters. In other words, we use the same set of values of all other parameters for the 28 observations (see the figure caption for details). This shows that the temperature of seed photons is the dominating factor affecting the total flux of the source when inverse Compton scattering is the key component of the radiation.

\begin{figure*}[!htbp]
\begin{center}
\includegraphics[width=2.0\columnwidth]{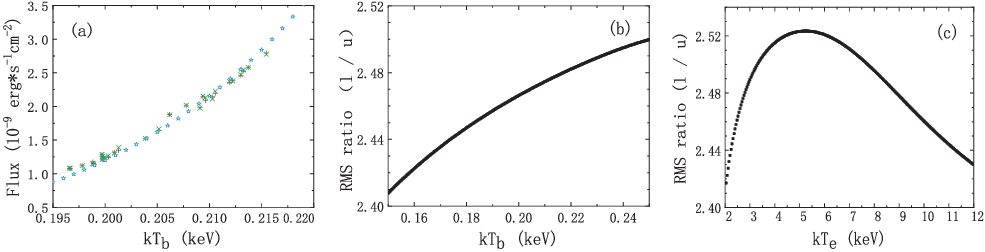}
  \caption{(a) The relations between the flux ($4\sim23\ \rm{keV}$) and the temperature of seed photons. The observed flux is marked by "plus", the model calculated flux is marked by "cross", and the flux marked by "star" is obtained using the same parameters, which are the central values of the parameters for the twin kHz QPOs with the frequencies ($\nu_{\rm l}=883.44\pm{1.10}\ {\rm Hz}, \nu_{\rm u}=1132.73\pm6.47\ {\rm Hz}$) in Table B.1 except the changed $kT_{\rm b}$. (b) The relations between the calculated rms ratio and the temperature of seed photons. (c) The relations between the calculated rms ratio and the temperature of electrons. Note: The data in panel (b) and (c) are obtained when the parameters in Table B.1 for the above twin kHz QPOs are kept unchanged except $kT_{\rm b}$ in panel (b) or are kept unchanged except $kT_{\rm e}$ in panel (c). }
  \label{fig12}
 \end{center}
\end{figure*}

In panel (b), rms ratio increases with increasing $k T_{\rm b}$ and the similar tendency can also be seen in panel (d) of Fig. 2. In panel (c), rms ratio first increases and then decreases with increasing $k T_{\rm e}$, which is similar to panel (c) of Fig. 3 for the higher $k T_{\rm e}$.
These similarities suggest that altering a single parameter has only a limited effect on the twin QPOs, while the optimal parameter
($kT_{\rm b}$) with stable results also influences the radiation.
However, comparisons become more challenging because Figs. 2 and 3 consider the properties of the entire kHz QPO sample, whereas Fig. 5 primarily uses parameters for only one pair of kHz QPOs. This indicates that different combination of the eight parameters will also lead to significant changes of the results.

\section{Discussion} \label{sec:S5}
\subsection{The physics behind the relations between the parameters}
The intricate relationship between kHz QPOs and the accretion-radiation processes within NS-LMXBs has been a subject of extensive exploration. KHz QPOs, characterized by their high-frequency nature, are considered as valuable probes into the underlying physics governing accretion onto NSs. Their frequency is intricately connected to the accretion rate, a crucial parameter dictating the innermost radius of the accretion disc, and the innermost radius of the accretion disc plays a pivotal role in determining the frequency of kHz QPOs. The higher the accretion rate, the smaller the innermost radius, leading to an increase in the frequency of these oscillations.
Several works, including those by \citet{Mendez1999,Mendez2001}, \citet{Straaten2002}, \citet{Done2007} and \citet{Altamirano2008}, have investigated the intricate connection between the innermost radius of the accretion disc and the X-ray luminosity of the source. Further exploration by \citet{Zhang2017}, \citet{Sanna2013} and \citet{Lyu2014} in the context of 4U 1636$-$53 has enriched our understanding of the interplay between spectral and timing properties.

Based on the simulation results presented in Sect. 3, we have illustrated the accretion and radiation processes for kHz QPOs in NS-LMXBs, depicting both low and high accretion rate scenarios in Fig. 6. The effect of changes in the corona's size relative to other parameters is not discussed here because no clear correlation has been found. Additionally, the transition between states, driven by the accumulation of material through the accretion process, requires a considerable amount of time. In the low accretion state, the accretion disc is truncated by a strong magnetic field with a larger magnetosphere, and the NS exhibits a lower surface temperature covered by a higher temperature corona. In contrast, under high accretion rate, the magnetosphere experiences more intense compression, leading to a reduction in the magnetospheric radius. Consequently, the accretion disc moves closer to the NS, and the NS, now with a higher surface temperature, is covered by a lower temperature corona.

 \begin{enumerate}
\item In panel (a) of Fig. 2, the flux of the source increases with the rising temperature of seed photons ($kT_{\rm b}$).  This correlation can be well described by an exponential function.
This suggests that, as the accretion rate increases,  the higher temperature of seed photons in the corona around the NS enhances the radiation of X-rays through Compton up-scattering. As illustrated in Eq. (1), the injection rate of seed photons per unit volume increases with the ascending $kT_{\rm b}$. The approximate relation (${\dot{n}}_{s\gamma}\sim E^2*e^{-\frac{E}{kT_{b}}}$) can be derived. Consequently, more seed photons are scattered in the corona, and their energy is boosted by the scattering process. This leads to an exponential rise in flux from the scattering with increasing $kT_{\rm b}$.
\item In contrast, the temperature of electrons in the corona decreases with the rising $kT_{\rm b}$, as depicted in panel (b) of Fig. 2, a phenomenon also discussed by Zhang et al. (2017). As explained earlier, the injection rate of seed photons increases with growing $kT_{\rm b}$, leading to a higher number density of photons in the corona. The increased photon density intensifies the cooling process of the corona due to the positive correlation between the cooling rate ($\int_{E_{\rm min}}^{E_{\rm max}}\frac{n_{\gamma{}0}{\sigma{}}_{\rm T}}{m_{\rm e}c}(4kT_{\rm e}-E)EdE$) and the photon density. In other words, the temperature ($kT_{\rm e}$) decreases when the rising Compton cooling rate per electron surpasses the heating rate, as per Eq. (4).
\end{enumerate}

Up to now, it is an open question how much depth MHD waves can be transported in the plasma accompanied with the strong magnetic field of a NS, and solving the problem will build up more knowledge of the relationships between the parameters. This endeavor will advance the understanding of the interaction between plasma and MHD waves, such as the propagation length of MHD waves, the frequencies of kHz QPOs and the heating rate of the plasma from MHD waves.

\begin{figure}[ht]
\begin{center}
\includegraphics[width=0.65\columnwidth]{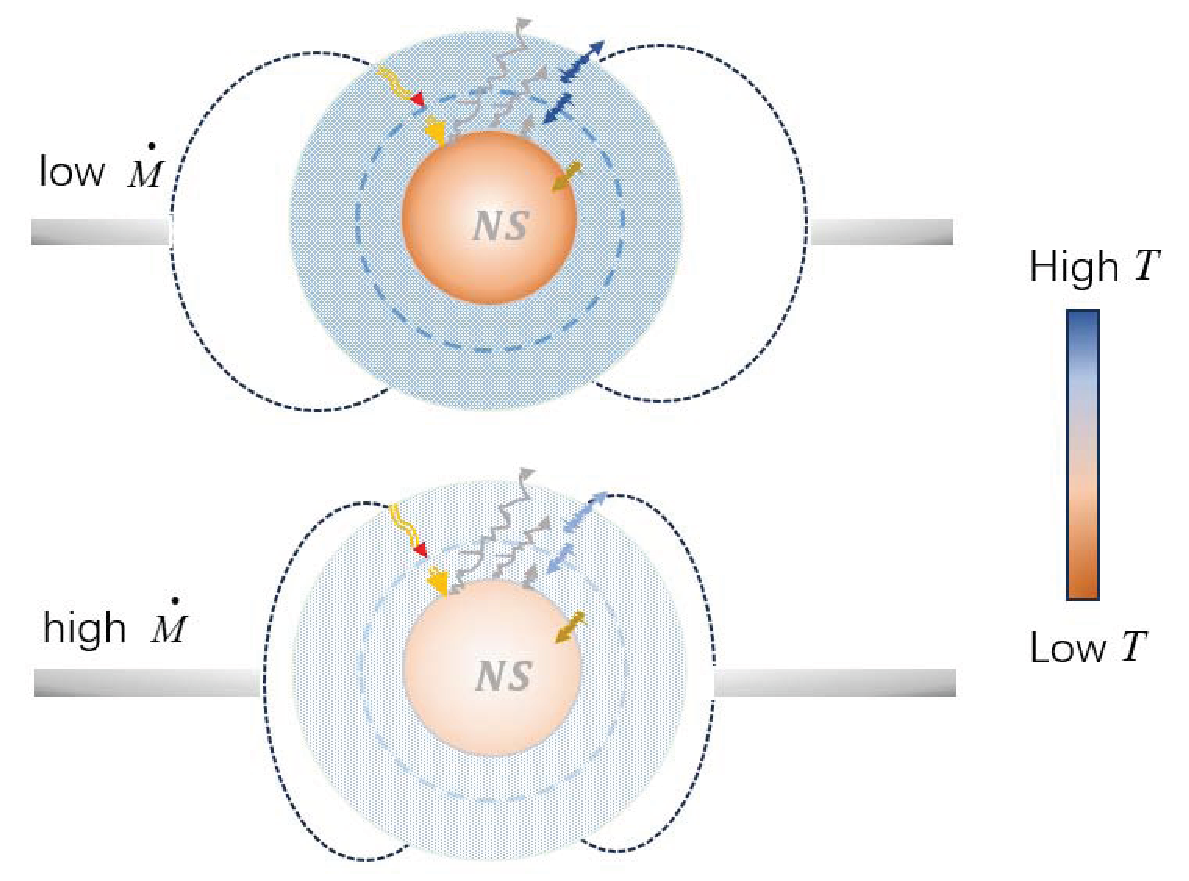}
  \caption{The sketch of the accretion and radiation process accompanied with kHz QPOs for the low or high accretion rate. Here $\dot{M}$ is the accretion rate of the disc and $T$ is the temperature of the corona.}
  \label{fig6}
 \end{center}
\end{figure}

\subsection{The possibly missing component in this model}
The parameters ($kT_{\rm e},\ kT_{\rm b},\ \tau,\ L, L_2$) are fitted by the spectra in the steady radiation, which are shown in Fig. 7. There are large discrepancies from zero for the residuals of the fitted spectra around $6.4\ \sim\ 7.0\ \rm{keV}$ in almost every panel. This likely indicates strong radiation from Fe k$\alpha$ or k$\beta$ lines. In addition, the large discrepancies from zero for the residuals of the fitted spectra in $E\  \lesssim 4\ \rm{keV}$ in most panels mean that the fit deviates from the observation from \citet{Zhang2017} because the component of the accretion disc and the directly escaped BB component in the energy region are ignored in our fitting, which can be seen in the fit of the characteristic spectrum in Fig. 3 of \citet{Zhang2017}. In Fig. 8, the ratios of the observational central values of flux per keV to the results in this model from the fits of the spectra are obtained. The hump around $6.4\ \sim\ 7.0\ \rm{keV}$ may originate from the deficiency of Fe emission lines in the fits. In most panels, the ratios deviate from unity and systematic errors from the deficiency of ignoring the Fe emission lines and the soft components exist in the fits.

In addition, the reduced $\chi^2$ is computed to estimate the quality of our fits in Fig. 9. The reduced $\chi^2$ can be obtained by $\frac{\chi^2}{N-N_0}$, where $N$, $N_0$ are the number of the data, the number of free parameters respectively. In addition, the general formula
$\chi^2 = \sum_{i=1}^{N} \chi_i^2 = \sum_{i=1}^{N} \frac{({{d_i}-{d_{i0}})}^2}{e_i^2}$ can be obtained according to the sum of every component ($\chi_i^2$) for the special energy photons, where ${d_i}$, ${e_i}$, ${d_{i0}}$ are the value of the observation, error and the theoretical value in our model, respectively.
As seen in most panels of Fig. 9 and Table 1, many components ($\chi_i^2$) of the reduced $\chi^2$ for the fit with the four
parameters slightly deviate from unity, which should originate from the large errors of the parameters ($kT_{\rm e},\ \tau,\ L$) for the fit of the four parameters. However, the fit with $L_{\rm 20}$ is very close to unity. The difference of $\chi_i^2$ between the two fits is
produced possibly by the different fitting methods. The central values of the fitted parameters in the first fit are substituted into
the second fit and thus it means that the central values of $kT_{\rm e},\ kT_{\rm b},\ \tau,\ L$ match the observation very well.
In addition, the component ($\chi_i^2$) in most observations (e.g. 30053-02-01-001, 40028-01-02-00, 60032-01-19-00, 60032-01-21-000,
93087-01-03-20) increases with the energy. This suggests that another component of the accretion system might be missing from the fit.
As proposed in our model, the accretion disc in 4U 1636$-$53 was omitted and it may be one of the reasons leading to the missing X-ray
component in 4U 1636$-$53. The humps in Fig. 9 (e.g. 10088-01-06-04, 10088-01-06-010, 50032-02-06-00, 60032-01-21-00, 60032-05-08-00)
around $6.4\ \sim\ 7.0\ \rm{keV}$ may also be produced by the omitted Fe lines in the fits. Accordingly, the fits may help identify
the missing components and the possible states of an XB related to the NS, the corona, the accretion disc, and other factors.
These components cannot be well constrained by RXTE data. In our future models, we will include the disc and reflection components and
test them using NICER data and the upcoming eXTP data.


\begin{figure*}[!htbp]
\begin{center}
\includegraphics[width=1.8\columnwidth]{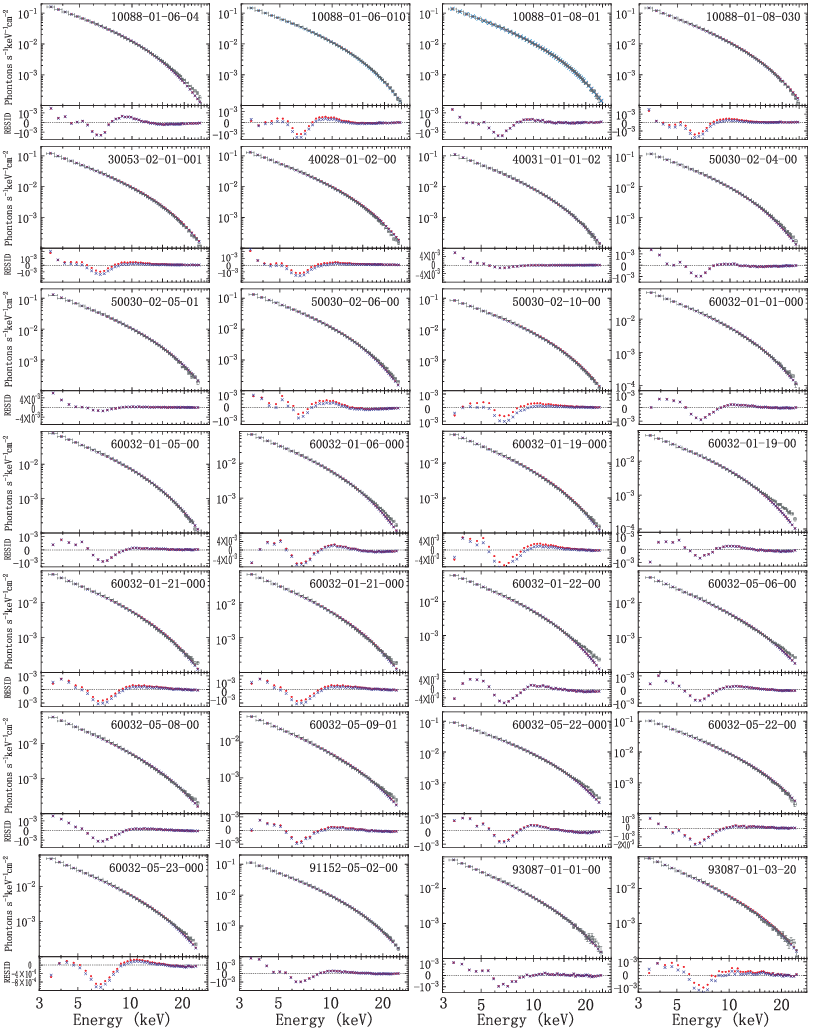}
  \caption{The fits and their residuals for the spectra in the 28 observations with the observation number in the upper right part of these panels in 4U 1636-53. The data described by the dots from the fits of $kT_{\rm e},\ kT_{\rm b},\ \tau,\ L$ and the data described by the crosses from the fits of $L_2$. All the units of the residuals are also $\rm {photons\ s^{-1}\ {keV}^{-1}\ {cm}^{-2}}$}.
  \label{fig7}
 \end{center}
\end{figure*}

\begin{figure*}[!htbp]
\begin{center}
\includegraphics[width=1.8\columnwidth]{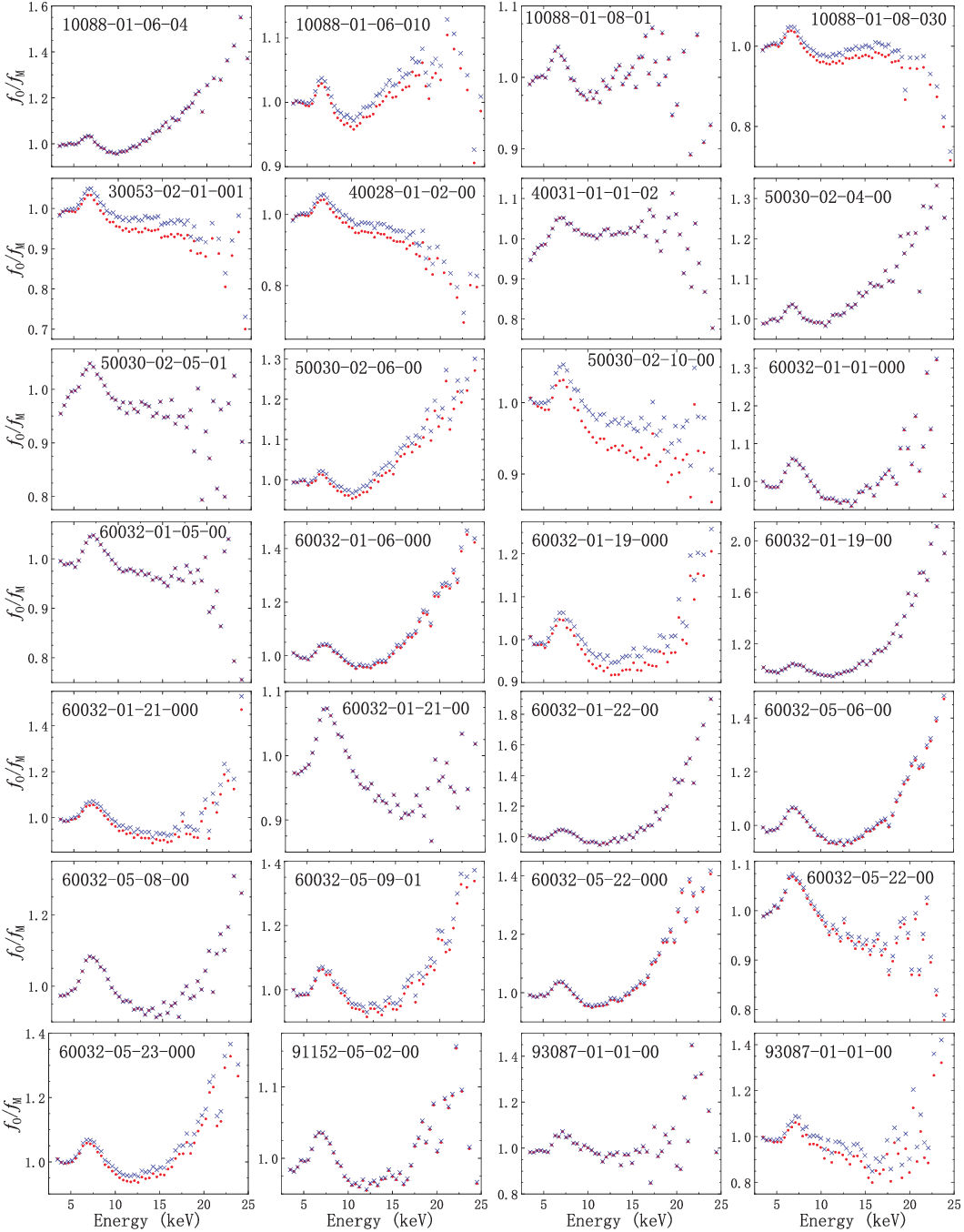}
  \caption{The ratios of the observational central values of flux per keV to the results in the model from the fits of the spectra in the 28 observations with the observation number in the upper right part of these panels in 4U 1636-53. The data described by the dots from the fits of $kT_{\rm e},\ kT_{\rm b},\ \tau,\ L$ and the data described by the crosses from the fits of $L_2$. }
  \label{fig8}
 \end{center}
\end{figure*}

\begin{figure*}[!htbp]
\begin{center}
\includegraphics[width=1.8\columnwidth]{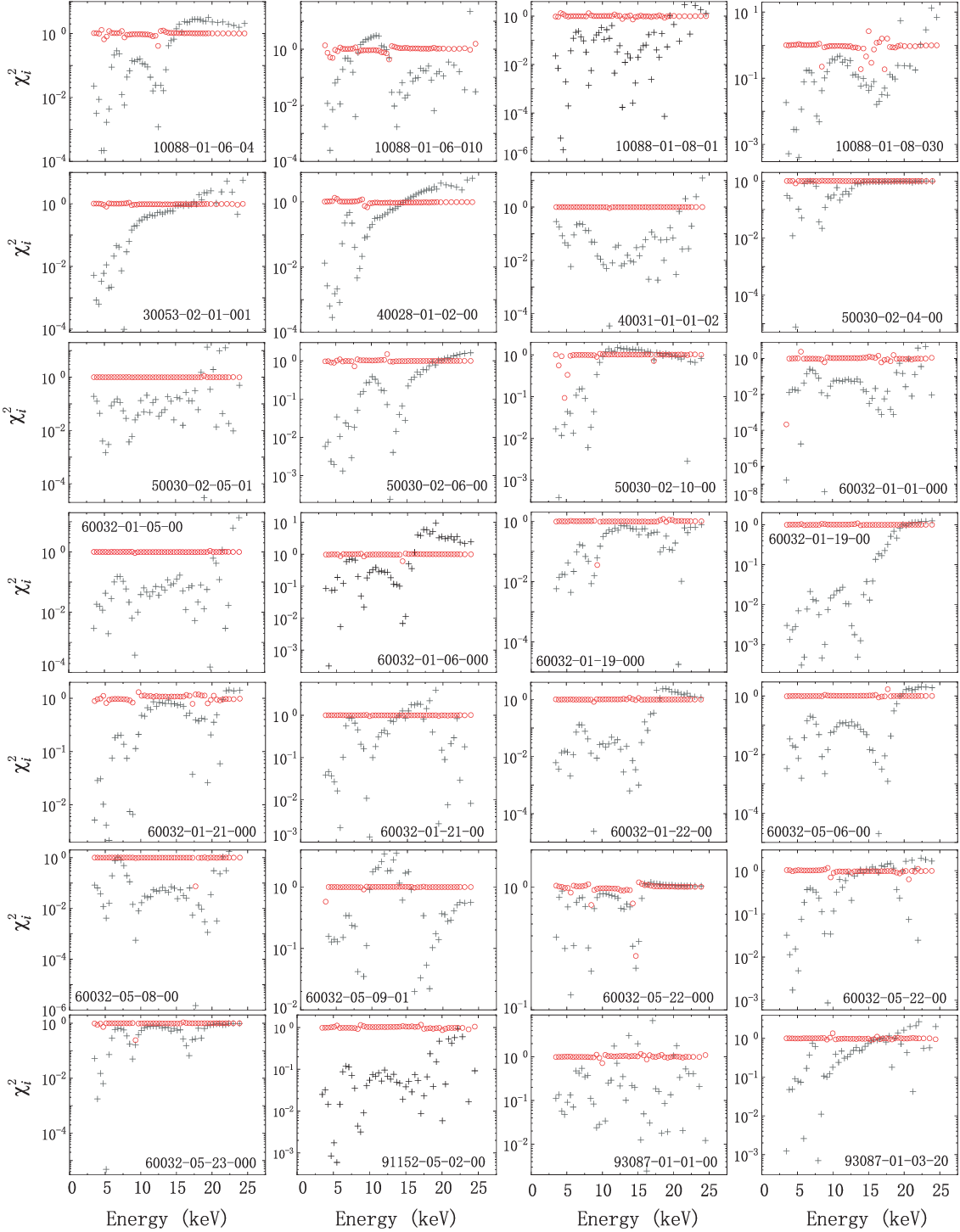}
  \caption{The component of $\chi_{i}^2$ from the fits of the spectra in the 28 observations with the observation number in the upper right part of these panels in 4U 1636-53. The data described by the plus sign from the fits of $kT_{\rm e},\ kT_{\rm b},\ \tau,\ L$ and the data described by circles from the fits of $L_2$. }
  \label{fig9}
 \end{center}
\end{figure*}

\subsection{The origin of QPOs}
Although the flare-related QPPs take place in the Sun but the QPOs take place in NS-LMXBs, they may have the same physical process but are shown in different wave bands, which should originate from different radiation mechanisms (e.g. gyrosynchrotron in the Sun and Compton up-scattering in NS-LMXBs). First, the arched magnetic field lines in the corona of the Sun are similar to the dipolar magnetic field lines of NSs. The magnetic field with arched field lines can make particles to be speeded up or down, which leads to that electromagnetic waves (e.g. X-rays) are produced. Then MHD waves lead to the oscillation of the magnetic field in a large region in a corona and some similar oscillation may also be produced in NS-LMXBs and the Sun. Finally, the dynamic interaction of waves and energetic particles appears in the region and variabilities of X-ray can be produced. According to the analysis, it can be guessed that the magnetic field with an arched field lines may be needed to produce QPOs.
However, the oscillations from different origins may lead to different types of QPOs. In the Sun, the short-period QPPs detected in radio/microwave emissions are considered to be related to the dynamic interaction of waves and energetic particles, while the long-period QPPs observed in white light, UV and EUV wavebands are usually thought to be associated with the dynamics of plasma. In brief, waves interacting with the plasmas and produced by dynamics may be an origin of QPOs. The reported arched quasi-periodic fast magnetoacoustic waves propagating away from the flare site (Liu et al. 2011) were found to be consistent with the periodicity of QPPs in the flare light curve, which may suggest an origin of QPPs. It help us to understand QPOs in LMXBs and other celestial bodies. However, the gravity magnetoacoustic waves as MHD waves may be a promising origin of QPOs due to the very strong gravity and magnetic field of a NS.

Here the frequencies of QPPs and QPOs ($\nu$) can be compared and estimated by using the expression of the frequency of Alf\'{v}en waves ($kV_{\rm A}\ \sim\ V_{\rm A}/r_{\rm c}$), which has the similar frequency to magnetoacoustic waves ($\omega^2\sim\frac{1}{2} k^2(V_{\rm A}^2+C_{\rm S}^2)$), where $k$, $V_{\rm A}$, $\omega$, $C_{\rm S}$, $r_{\rm c}$ are the wave number, the Alf\'{v}en speed, the angular frequency and the sonic speed of a MHD wave, the characteristic length of the magnetic field lines, respectively. The frequency ratio of kHz QPOs to QPPs ($10^2 \sim 10^6$) is consistent with the ratio of $kV_{\rm A}$ ($10^2 \sim 10^8$) that is obtained by substituting the relations ($V_{\rm A}=B/\sqrt{\mu \rho}$, $\rho \sim n $) and the characteristic parameters in Table 1 into the frequency expression, where $B$, $\mu$, $\rho$, $n$ are the characteristic magnetic field, magnetic conductivity, density and number density of plasma, respectively. This ratio supports the connection between the kHz QPOs and the frequencies of QPPs.
It's worth noting that this correlation does not exclude the possibility of oscillations in the plasma being generated by other physical processes, such as rotation, magnetic reconnection, or precession, within the mechanism of QPOs.

\begin{table}[htbp!]
\label{appB}
\caption{The comparison of the characteristic parameters.}
\centering
\small
\begin{tabular}{l|c|c}
\hline   \hline
&{The corona loop of the Sun} &{The corona around the NS}\\
  \hline
{$r_{\rm c}\ ({\rm {cm}}) $}&{$10^9\sim10^{10}$}\tablefootmark{, a}& $10^6\sim10^7$\tablefootmark{, b}\\
    \hline
$ n\ ({\rm{cm}^{-3}}) $ & $ 10^8\sim10^{10}\tablefootmark{, a} $ & $ 10^{18}\sim10^{20}  $\tablefootmark{, c} \\
    \hline
$ B\ (\rm G) $ & $ 10\sim100$\tablefootmark{a} & $ 10^8\sim 10^9 $\tablefootmark{, d} \\
   \hline
$ \nu\ (\rm {Hz}) $ & $ 0.001 \sim 1$\tablefootmark{a} & $10^2 \sim 10^3$\tablefootmark{, d} \\
   \hline
\end{tabular}
\tablefoot{
This table lists the characteristic parameters of the Corona loop of the Sun and the corona around the NS in a LMXB. The number density of electrons is considered as the indicative number density of the plasma in the corona around the NS in a LMXBs.\\
\tablefoottext{a}{\citet{Zimovets2021}}; \tablefoottext{b}{\citet{Karpouzas2020}}; \tablefoottext{c}{\citet{Zhang2017}}; \tablefoottext{d}{\citet{Wijnands2006}}.
}
\end{table}

 According to this work, lower QPOs are produced in the whole corona but upper QPOs are produced in the outmost layer of the corona. The seed photons producing the upper QPOs are from the radiation in the inner layer and harder than the BB seed photons from the NS on average. It was also reported that scattering in the upper layers of a NS corona produced hardening of the spectra (Lapidus et al. 1986, London et al. 1986). Thus the average energy of the last photons producing the upper QPOs should be higher than that of the photons for the lower QPOs. It may be the key reason that the lower QPOs have different rms from the upper QPOs. In addition, most of the soft photons from the inner layer will travel a longer distance than hard photons and thus many soft photons will take a longer time in order to arrive at the surface of the corona, which can partially explain the time-lag between soft and hard X-ray photons. These two characteristics are the key differences between the upper QPOs and the lower QPOs, which will be studied in a next work. Finally, a single MHD wave, which may be an origin of the single kHz QPO \citep{Shi2014}, can also make the physical quantities perturbed and leads to the kHz X-ray variability. Therefore the radiation mechanism of the single kHz QPOs can be further investigated based on the findings of this study.

\section{Conclusion} \label{sec:S6}
In this work we combine the dynamic mechanism and radiation process of the twin kHz QPOs described by Compton up-scattering, in which 28 twin kHz QPOs in 4U 1636$-$53 are considered. The parameters in the corona around the NS in the LMXB for the 28 twin kHz QPOs are obtained by Monte Carlo method. Our conclusions can be summarized as follows.
\begin{enumerate}
\item The interaction between MHD waves and plasma in one part of a corona can be explored by layering the corona. QPOs can be considered as a disturbance superimposed on top of the radiation of the radiation field in a quasi-steady state.
\item Twin disturbances from twin MHD waves produced at the innermost radius of an accretion disc can be considered as the origin of QPOs. Seed photons may be transported through a high temperature corona and scattered by Compton up-scattering mechanism. Finally, the variability of radiation photons with the frequencies of the twin MHD waves will lead to the observed twin kHz QPOs.
\item According to this model, the calculated flux of the source increases sensitively with the increasing temperature of seed photons. In addition, rms ratio of lower QPOs to that of upper QPOs increases slightly with the increasing temperature of seed photons.
\item Considering the chosen best parameters, the weak relations between $k T_{\rm e}$ and $f$, $f_{\rm L}/f_{\rm U}$,
${\rm RMSL}/{\rm RMSU}$ are obtained. With the increasing temperature of electrons, all the three parameters decreases, which shows that the observational characteristics of QPOs are possibly closely linked to the accretion state of the source.
In addition, the negative correlations are also shown between $k T_{\rm e}$ and $k T_{\rm b}$.
\end{enumerate}

 In summary, twin kHz QPOs can be considered to originate from the oscillating physical parameters of the plasma in a whole corona and in a part of the corona. Twin kHz QPOs may be a key tool to decide the innermost radius of an accretion disc, study the characteristics of state in LMXBs, test effect of general relativity, estimate parameters of compact stars and explore the new MHD waves (gravity magnetoacoustic waves).

\begin{acknowledgements}
    We thank Philipp Podsiadlowski, Mariano M{\'e}ndez, Konstantinos Karpouzas and Adam Ingram for the helpful discussion. This work was supported by the Hainan Provincial Natural Science Foundation of China under grant Nos. 122RC546, 124CXTD422, the National Natural Science Foundation of China under grant Nos. 12373043, 12063001, 11563003, 12333007, 12027803, 12041301, and 12121003, and the National Key Research and Development Program of China (2021YFA0718500). GB acknowledge the science research grants from the China Manned Space Project.
\end{acknowledgements}

\nocite{*}
\bibliographystyle{aa}
\bibliography{aanda}

\begin{appendix}
\onecolumn
\section{The formula of Compton scattering}
\label{appA}
The parameter ($\varepsilon$) in Kompaneets equations can be expressed as,
 \begin{equation}
  \label{eq1}
\varepsilon=\left\{
\begin{array}{rcl}
1                                                       &      {\rm for}          &   E <0.1\ m_{\rm e}c^2,\\
(1-\frac{E}{m_{\rm e}c^2}) /0.9       &      {\rm for}          &  0.1\ m_{\rm e}c^2<E<m_{\rm e}c^2,\\
0                                                      &       {\rm for}         &   E > m_{\rm e}c^2 .\\
\end{array}
\right.
\end{equation}

The Klein-Nishina cross section is,
 \begin{equation}
   \label{eq2}
   \begin{aligned}
 \sigma_{\rm KN}=&\frac{3}{4}{\sigma{}}_{\rm T}[\frac{1+x}{x^3}(\frac{2x\left(1+x\right)}{1+2x}-\ln{\left(1+2x\right))}
 &+\frac{\ln{\left(1+2x\right)}}{2x}-\frac{1+3x}{{(1+2x)}^2}],
\end{aligned}
\end{equation}
where $x=\frac{E}{m_{\rm e}c^2}$.

\section{The parameters in the states of 4U 1636$-$53 with 28 twin kHz QPOs}
\label{appB}
\begin{sidewaystable}[]
\caption{The frequencies of twin kHz QPOs and the best parameters in 4U 1636--53.}
\setlength{\tabcolsep}{2.5mm}{
\begin{tabular}{lc|ccccc|cc|ccr}
\\   \hline
  \hline\\
  \multicolumn{2}{c|}{Input parameters}&\multicolumn{5}{c|}{The first fitted parameters}  & \multicolumn{2}{c|}{The second fitted parameter} &\multicolumn{3}{c}{Characteristic values }\\
  \hline
{$f_{\rm l}\ ({\rm Hz})$}&{$f_{\rm u}\ ({\rm Hz})$} & $\tau$ & $L\ ({\rm km})$ &$k T_{\rm e}\ ({\rm keV})$& $k T_{\rm b}\ ({\rm keV})$ &  $\mathcal{X}^2/d.o.f.$ & $L_2\ ({\rm km})$ &  $\mathcal{X}^2/d.o.f.$  & $\eta$ & $\Delta {\dot{H}_1}$ & $\Delta {\dot{H}_2}$\\
    \hline
$ 558.15  \pm 6.92 $ & $ 879.25_{-5.61}^{+5.49} $ & $ 8.4  \pm 0.5  $ & $ 9.1  \pm 2.2  $ & $ 3.6  \pm 0.3  $ & $ 0.1966  \pm 0.0002  $ & 36 / 44 & $ 9.08  \pm 0.01  $ & 47 / 47 & 1.0  & 9.69  & 16.20  \\
$ 574.42  \pm 5.43 $ & $ 884.05_{-4.40}^{+4.52} $ & $ 9.4  \pm 0.1  $ & $ 3.7  \pm 0.1  $ & $ 3.16  \pm 0.01  $ & $ 0.2062  \pm 0.0001  $ & 38 / 44 & $ 3.670  \pm 0.002  $ & 47 / 47 & 0.0  & 4.05  & 13.70  \\
$ 575.87  \pm 2.74  $ & $ 909.49  \pm 3.83 $ & $ 8.9  \pm 0.2  $ & $ 12.0  \pm 0.4  $ & $ 3.33  \pm 0.07  $ & $ 0.19971  \pm 0.00004  $ & 25 / 44 & $ 11.880  \pm 0.009  $ & 47 / 47 & 1.0  & 12.86  & 27.65  \\
$ 590.01  \pm 6.37 $ & $ 867.78  \pm 3.78 $ & $ 8.2  \pm 0.2  $ & $ 9.8  \pm 1.8  $ & $ 3.63  \pm 0.07  $ & $ 0.1966  \pm 0.0001  $ & 20 / 44 & $ 9.722  \pm 0.007  $ & 49 / 47 & 1.0  & 14.73  & 25.04  \\
$ 612.14  \pm 5.04 $ & $ 906.11  \pm 4.47 $ & $ 8.3  \pm 0.4  $ & $ 5.7  \pm 0.3  $ & $ 3.46  \pm 0.2  $ & $ 0.1997  \pm 0.0001  $ & 24 / 45 & $ 5.63  \pm 0.00  $ & 49 / 48 & 1.0  & 12.77  & 15.01  \\
$ 637.41  \pm 4.86 $ & $ 930.24_{-12.55}^{+12.70} $ & $ 10.1  \pm 0.1  $ & $ 4.0  \pm 0.9  $ & $ 2.89  \pm 0.07  $ & $ 0.1988  \pm 0.0001  $ & 14 / 44 & $ 3.953  \pm 0.002  $ & 48 / 47 & 0.5  & 11.94  & 13.62  \\
$ 640.23  \pm 2.72 $ & $ 964.57  \pm 5.50  $ & $ 10.0  \pm 0.4  $ & $ 6.0  \pm 0.1  $ & $ 2.9  \pm 0.1  $ & $ 0.20089  \pm 0.00004  $ & 77 / 44 & $ 5.914  \pm 0.005  $ & 47 / 47 & 0.5  & 12.96  & 14.91  \\
$ 649.55  \pm 4.28 $ & $ 936.25  \pm 10.28 $ & $ 8.3  \pm 0.6  $ & $ 2.6  \pm 0.3  $ & $ 3.5  \pm 0.3  $ & $ 0.1978  \pm 0.0004  $ & 36 / 44 & $ 2.5603  \pm 0.0001  $ & 56 / 47 & 0.4  & 13.08  & 13.62  \\
$ 650.68  \pm 4.23 $ & $ 991.97_{-11.29}^{+8.71} $ & $ 8.6  \pm 0.5  $ & $ 6.9  \pm 0.5  $ & $ 3.3  \pm 0.2  $ & $ 0.2078  \pm 0.0003  $ & 32 / 44 & $ 6.841  \pm 0.007  $ & 46 / 47 & 0.1  & 10.14  & 12.14  \\
$ 653.58_{-3.82}^{+3.66} $ & $ 984.64  \pm 5.69 $ & $ 8.4  \pm 0.7  $ & $ 7.9  \pm 0.1  $ & $ 3.5  \pm 0.4  $ & $ 0.2013  \pm 0.0002  $ & 30 / 45 & $ 7.694  \pm 0.018  $ & 49 / 48 & 0.0  & 16.19  & 22.56  \\
$ 659.10  \pm 2.12 $ & $ 975.20  \pm 4.12 $ & $ 9.6  \pm 0.8  $ & $ 5.4  \pm 0.4  $ & $ 2.9  \pm 0.3  $ & $ 0.1989  \pm 0.0002  $ & 26 / 44 & $ 5.385  \pm 0.001  $ & 48 / 47 & 0.6  & 17.03  & 13.25  \\
$ 675.51  \pm 1.42 $ & $ 970.16  \pm 4.83 $ & $ 9.2  \pm 0.8  $ & $ 6.4  \pm 0.9  $ & $ 3.1  \pm 0.3  $ & $ 0.2094  \pm 0.0001  $ & 6 / 45 & $ 6.392  \pm 0.003  $ & 49 / 48 & 0.0  & 10.88  & 12.45  \\
$ 689.14  \pm 42.36 $ & $ 1022.49  \pm 42.36 $ & $ 9.2  \pm 0.9  $ & $ 2.8  \pm 0.6  $ & $ 3.1  \pm 0.3  $ & $ 0.2000  \pm 0.0001  $ & 23 / 44 & $ 2.751  \pm 0.003  $ & 50 / 47 & 0.1  & 14.60  & 11.47  \\
$ 696.99  \pm 1.01 $ & $ 1007.83  \pm 9.49  $ & $ 8.6  \pm 0.3  $ & $ 4.8  \pm 0.8  $ & $ 3.3  \pm 0.1  $ & $ 0.1998  \pm 0.0001  $ & 30 / 44 & $ 4.7770  \pm 0.0003  $ & 48 / 47 & 0.5  & 19.93  & 12.71  \\
$ 700.10  \pm 2.72 $ & $ 1005.88  \pm 7.68 $ & $ 9.5  \pm 0.7  $ & $ 8.4  \pm 0.7  $ & $ 3.0  \pm 0.3  $ & $ 0.2003  \pm 0.0002  $ & 15 / 44 & $ 8.332  \pm 0.006  $ & 47 / 47 & 1.0  & 24.10  & 14.50  \\
$ 718.56 _{-1.78}^{+1.81} $ & $ 988.76  \pm 12.25 $ & $ 9.5  \pm 0.2  $ & $ 5.3  \pm 0.6  $ & $ 2.99  \pm 0.07  $ & $ 0.19973  \pm 0.00006  $ & 13 / 44 & $ 5.256  \pm 0.003  $ & 49 / 47 & 0.4  & 20.84  & 14.45  \\
$ 770.02  \pm 42.55 $ & $ 1092.21  \pm 42.55 $ & $ 9.7  \pm 0.4  $ & $ 5.0  \pm 0.9  $ & $ 2.9  \pm 0.1  $ & $ 0.2039  \pm 0.0003  $ & 24 / 44 & $ 4.9465  \pm 0.0003  $ & 48 / 47 & 0.2  & 18.86  & 10.09  \\
$ 772.80  \pm 1.33 $ & $ 1070.88  \pm 7.04 $ & $ 9.9  \pm 0.4  $ & $ 8.2  \pm 0.3  $ & $ 2.9  \pm 0.1  $ & $ 0.2052  \pm 0.0001  $ & 36 / 44 & $ 8.115  \pm 0.005  $ & 46 / 47 & 0.1  & 21.47  & 13.77  \\
$ 877.07  \pm 29.23 $ & $ 1126.12  \pm 29.23 $ & $ 10.0  \pm 0.2  $ & $ 7.5  \pm 0.1  $ & $ 2.75  \pm 0.06  $ & $ 0.2137  \pm 0.0002  $ & 51 / 48 & $ 7.473  \pm 0.004  $ & 52 / 51 & 0.0  & 12.04  & 7.86  \\
$ 883.19  \pm 3.67 $ & $ 1187.03  \pm 6.72 $ & $ 9.2  \pm 0.3  $ & $ 3.9  \pm 0.2  $ & $ 3.0  \pm 0.1  $ & $ 0.2106  \pm 0.0001  $ & 59 / 48 & $ 3.871  \pm 0.001  $ & 51 / 51 & 0.0  & 9.53  & 4.88  \\
$ 883.44  \pm 1.10 $ & $ 1132.73  \pm 6.47 $ & $ 10.2  \pm 0.6  $ & $ 6.4  \pm 0.4  $ & $ 2.7  \pm 0.1  $ & $ 0.2130  \pm 0.0005  $ & 21 / 44 & $ 6.351  \pm 0.003  $ & 48 / 47 & 0.0  & 10.79  & 7.05  \\
$ 913.75  \pm 40.91 $ & $ 1161.56  \pm 40.91 $ & $ 9.2  \pm 0.1  $ & $ 8.7  \pm 0.6  $ & $ 2.94  \pm 0.07  $ & $ 0.2133  \pm 0.0003  $ & 36 /  48 & $ 8.644  \pm 0.005  $ & 50 / 51 & 0.0  & 12.12  & 7.97  \\
$ 913.98  \pm 2.97  $ & $ 1223.44  \pm 6.32 $ & $ 9.4  \pm 0.1  $ & $ 4.8  \pm 1.5  $ & $ 2.98  \pm 0.03  $ & $ 0.20961  \pm 0.00003  $ & 43 /  48 & $ 4.713  \pm 0.001  $ & 51 / 51 & 0.0  & 9.07  & 4.40  \\
$ 922.59_{-2.13}^{+1.62} $ & $ 1164.83  \pm 5.82 $ & $ 8.9  \pm 0.4  $ & $ 5.3  \pm 0.8  $ & $ 3.0  \pm 0.1  $ & $ 0.2119  \pm 0.0001  $ & 44 / 44 & $ 5.2900  \pm 0.0004  $ & 48 / 47 & 0.0  & 8.93  & 6.17  \\
$ 923.31  \pm 3.91 $ & $ 1193.66_{-4.91}^{+4.78} $ & $ 10.3  \pm 0.3  $ & $ 6.9  \pm 0.4  $ & $ 2.64  \pm 0.05  $ & $ 0.2154  \pm 0.0002  $ & 48 / 48 & $ 6.869  \pm 0.003  $ & 52 / 51 & 0.0  & 8.46  & 4.90  \\
$ 925.03_{-4.43}^{+4.44} $ & $ 1191.40  \pm 2.57 $ & $ 9.6  \pm 0.3  $ & $ 6.3  \pm 0.6  $ & $ 2.82  \pm 0.09  $ & $ 0.2122  \pm 0.0002  $ & 14 / 48 & $ 6.219  \pm 0.002  $ & 52 / 51 & 0.0  & 9.20  & 5.57  \\
$ 932.44_{-5.14}^{+5.28} $ & $ 1191.81  \pm 5.26 $ & $ 9.96  \pm 0.06  $ & $ 7.1  \pm 0.1  $ & $ 2.73  \pm 0.01  $ & $ 0.21029  \pm 0.00006  $ & 32 /  44 & $ 7.0306  \pm 0.0001  $ & 48 / 47 & 0.0  & 10.77  & 6.88  \\
$ 948.89_{-14.11}^{+11.18} $ & $ 1197.21_{-3.77}^{+3.68} $ & $ 9.1  \pm 0.2  $ & $ 9.9  \pm 0.7  $ & $ 2.91  \pm 0.07  $ & $ 0.2091  \pm 0.0002  $ & 20 / 44 & $ 9.8445  \pm 0.0001  $ & 48 / 47 & 0.1  & 15.10  & 10.11  \\
   \hline
  \hline
\end{tabular}
}
\tablefoot{
This table lists the frequencies of 28 twin kHz QPOs and the best parameters in 4U 1636--53 for every group of twin QPOs. The first two parameters ($f_{\rm l}$ and $f_{\rm u}$) in the first column are from the observation and the values of other parameters are obtained from this model.\\}
\end{sidewaystable}
\end{appendix}

\clearpage
 \newpage

\end{document}